\documentclass[5p,times]{elsarticle}

\usepackage{amssymb}

\usepackage{url}

\usepackage[shortcuts]{extdash}

\usepackage{multirow}

\usepackage{lineno}

\usepackage{listings}
\usepackage{xcolor}

\definecolor{codegreen}{rgb}{0,0.6,0}
\definecolor{codegray}{rgb}{0.5,0.5,0.5}
\definecolor{codepurple}{rgb}{0.58,0,0.82}
\definecolor{backcolour}{rgb}{0.95,0.95,0.92}

\lstdefinestyle{mystyle}{
    backgroundcolor=\color{backcolour},   
    commentstyle=\color{codegreen},
    keywordstyle=\color{magenta},
    numberstyle=\tiny\color{codegray},
    stringstyle=\color{codepurple},
    basicstyle=\ttfamily\footnotesize,
    breakatwhitespace=false,         
    breaklines=true,                 
    captionpos=b,                    
    keepspaces=true,                 
    numbers=left,                    
    numbersep=5pt,                  
    showspaces=false,                
    showstringspaces=false,
    showtabs=false,                  
    tabsize=2
}
\journal{FGCS}

\begin{document}

\begin{frontmatter}

\title{Enhancing iteration performance on distributed task-based workflows}

\author{
  Alex Barcelo\\
  \emph{Barcelona Supercomputing Center}
}
\address{alex.barcelo@bsc.es}

\author{
  Anna Queralt\\
  \emph{Universitat Polit\`ecnica de Catalunya\\
  Barcelona Supercomputing Center}
}
\address{anna.queralt@upc.edu}

\author{
  Toni Cortes\\
  \emph{Universitat Polit\`ecnica de Catalunya\\
  Barcelona Supercomputing Center}
}
\address{toni.cortes@upc.edu}

\begin{abstract}

Task-based programming models have proven to be a robust and versatile way to approach development of applications for distributed environments. They provide natural programming patterns with high performance. However, execution on this paradigm can be very sensitive to granularity --i.e., the quantity and execution length of tasks. Granularity is often linked with the block size of the data, and finding the optimal block size has several challenges, as it requires inner knowledge of the computing environment.

Our proposal is to supplement the task-based programming model with a new mechanism --our \emph{SplIter} proposal. At its core, the \emph{SplIter} provides a transparent way to \emph{split} a collection into \emph{partitions} (logical groups of blocks, obtained without any transfers nor data rearrangement), which can then be \emph{iterated}. Tasks are linked to those partitions, which means that \emph{SplIter} breaks the dependency between block size and task granularity. 

The evaluation shows that the \emph{SplIter} is able to achieve performance improvements of over one order of magnitude when compared to the baseline, and it is either competitive or strictly better (depending on application characteristics) to the competitor alternative. We have chosen different applications covering a wide variety of scenarios; those applications are representatives of a broader set of applications and domains.
The changes required in the source code of a task-based application are minimal, preserving the high programmability of the programming model. Two different state-of-the-art task-based frameworks have been evaluated for all the applications: COMPSs and Dask, showing that the \emph{SplIter} can be effectively used within different frameworks.

\end{abstract}


\end{frontmatter}



\section{Introduction}

We have been witnessing a sustained growth on the available computing capabilities for decades now. 
Nodes keep becoming more powerful (with faster microprocessors and bigger memory). Simultaneously, clusters keep increasing in size (number of nodes, storage). In High Performance Computing environments, this increase on computation resources has enabled new and improved applications. For example, data analytics algorithms can be executed onto datasets that were unmanageable years ago and machine learning applications are now run faster and more accurately.

At a high abstraction level we can say that parallelism is the main factor that is able to sustain all these improvements. The implementation of algorithms consists of the iteration on datasets and the execution of operations unto them. The execution will then be parallelized within and between nodes. This division of work goes along the distribution of data --i.e. the blocking procedure, by which a dataset is divided into blocks.

Being able to perform parallel and distributed execution on distributed data is complex; two branches of programming models aim to address that with very different approaches: map-reduce\cite{dean2008mapreduce,zaharia2016apache} and task-based\cite{matthew_rocklin-proc-scipy-2015,compss}. Those two paradigms have been coexisting for a while, each one with their own strengths. Both paradigms make use of blocks for achieving a distributed execution. However, their approach differs: the map-reduce paradigm abstracts the notion of blocks and iteration into \emph{map} and \emph{reduce} operations, while we can find explicit operations and blocking directives in task-based programming models. The abstractions provided by  map-reduce allow the framework implementation to perform blocking implicitly as well as to introduce transparent iteration optimizations.

The strengths of task-based programming models are their high programmability and their flexibility\cite{perez2008dependency,kaiser2014hpx}. Implementation of scientific algorithms with a task-based approach causes much less friction and results in a more agile cycle of design, implementation, and testing. However, when developing under these kinds of programming models, there is a parameter one must be wary about: the block size. Decreasing the block size implies more potential parallelism but also results in an increase of the number of blocks --resulting in a higher number of faster tasks. Consequently, the scheduler stress rises and the runtime invocation overhead increases. This is a limitation of the programming model: the granularity of the computation is linked to the granularity of the data. The optimal value will depend on the computing capabilities of the environment as well as on the application behavior and its iteration and data access patterns. Having to set up this value is a burden on the developer that also reduces the performance portability.

Our proposal is to solve the granularity issue by providing a mechanism that yields partitions of the data: \emph{SplIter}, which can be included within the existing iterations of applications. The main idea is that the \emph{partitions} encompass multiple \emph{blocks} (in a logical way). By using these partitions as inputs for the tasks, we successfully decouple the task granularity from the block granularity. \emph{SplIter} is able to automatically leverage runtime information; this information includes computation capabilities of the environment --such as number of nodes and number of cores-- as well as location of the data blocks --information that can be used to exploit data locality and avoid data transfers. The application will still use blocks (with a certain block size); however, by following our proposal, the performance sensitivity to the block size will be greatly reduced, as we will later show.


\section{Related work}

\label{sec:relatedwork}

\begin{figure}
    \centering
    \includegraphics[width=0.95\columnwidth]{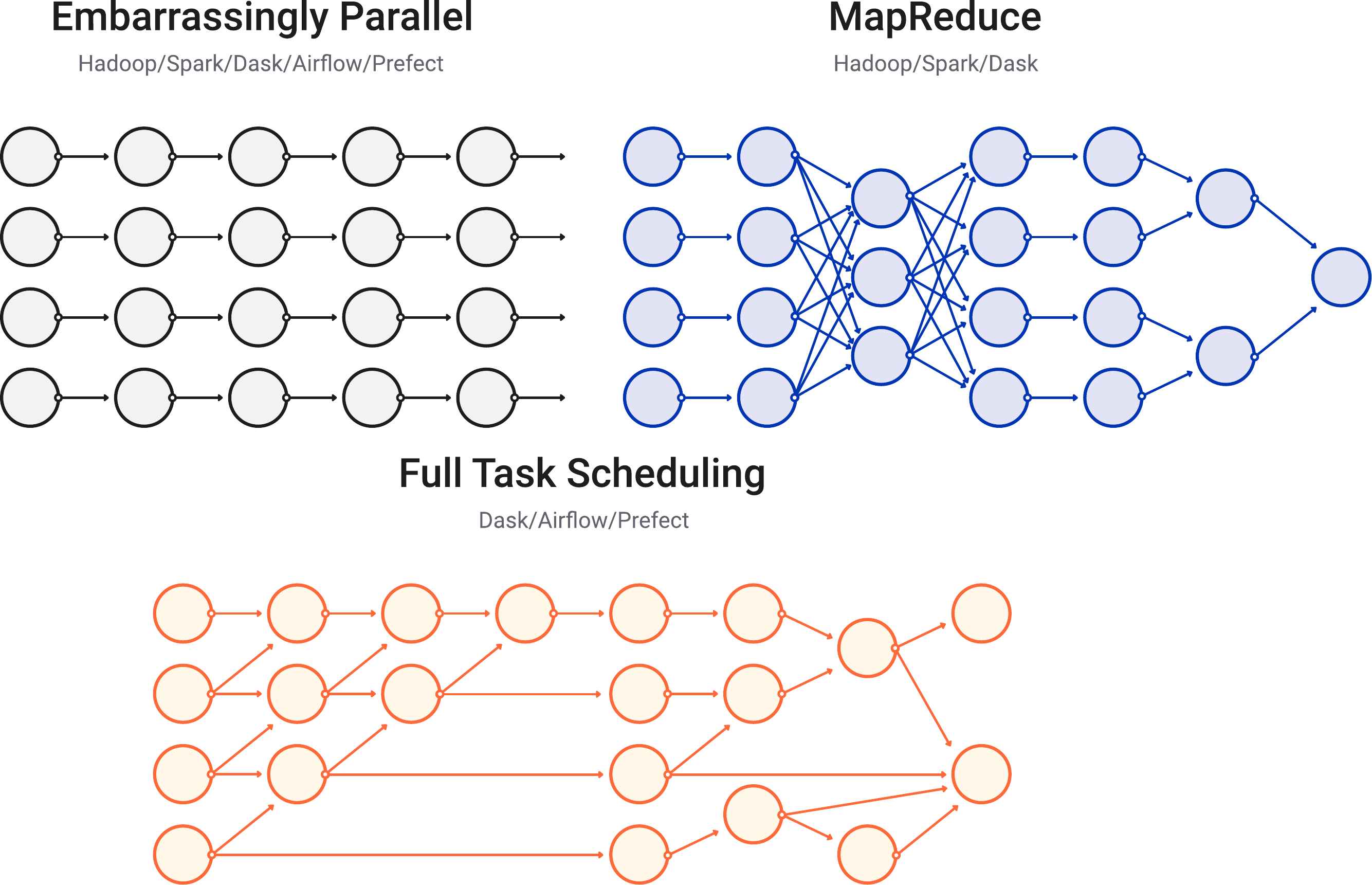}
    \caption{Different methods for task scheduling, from Dask documentation\cite{daskweb}}
    \label{fig:task-scheduling-methods}
\end{figure}

One primary programming model that tackles the topic of iteration on distributed datasets is MapReduce\cite{dean2008mapreduce}. The abstractions of the programming model --which effectively hide the iteration from the developer-- are mechanisms that the framework uses to achieve good performance. An application following the map-reduce programming model consists of \emph{map} tasks (which transform blocks, and are applied to all input blocks of the input) and \emph{reduce} operations (which aggregate multiple input blocks into single outputs). This general process is depicted in Figure~\ref{fig:task-scheduling-methods}. The programming model hides the blocking and iteration, which is managed transparently by the framework, allowing for internal optimizations --even if the iteration is not present in the application code, the framework implementation does iterate during runtime. The inherent limitation of this programming model is that it cannot be applied to all kind of applications, as not all algorithms can feasibly be translated into \emph{map} and \emph{reduce} tasks. Our proposal is to maintain the iterations at the programming model level while simultaneously providing enhanced iteration code structures.

Spark\cite{zaharia2016apache} is a widely used software stack that aims to be a unified engine for large-scale data analytics. It draws inspiration of the MapReduce programming model but aims to be more flexible and improve its performance. The distributed data structure used by this programming model is the Resilient Distributed Dataset or RDD\cite{zaharia2012resilient}, which carries the burden of blocking and distributing while leveraging performance optimizations within the framework implementation. 
However, using Spark requires to adapt (and rewrite) existing algorithms into using the explicit primitives and structures provided by the framework\cite{conejero2018task}. 
With its mechanisms, Spark does address the issues of blocking and data distribution, but restricts the application developer to their programming primitives as well as to the data structures provided by the framework --namely, the  Resilient Distributed Dataset or its derived abstractions, which requires the application developer to adapt their code to the new data interface. We will instead focus on addressing the iterations --the explicit iterations at the programming model level-- with a minimal impact on application programmability.

A different software stack, one that follows a task-based full scheduling paradigm but shares common goals with the previous one is Dask\cite{matthew_rocklin-proc-scipy-2015}. It is a ``flexible library for parallel computing in Python''. It has a very flat learning curve for Python developers, as it draws inspiration from commonly used data structures (such as NumPy \emph{Arrays} and Pandas \emph{DataFrames}). With those data structures in mind, Dask provides their distributed counterparts (e.g. \emph{Dask Array} and \emph{Dask DataFrame}). Its flexible task scheduling mechanism is shown in Figure~\ref{fig:task-scheduling-methods}. Dask library provides a lot of functionality out-of-the-box (for instance, Dask arrays support most of the NumPy interface, which helps flatten the learning curve). It is worth mentioning that Dask is able to exploit data locality by leveraging the Python memory space of its workers.

COMPSs\cite{compss} is another a task-based framework, with a robust and complete scheduler with support for complex interaction of data dependencies. The \texttt{dislib}\cite{dislib} library provides a distributed data structure along data analytics algorithms implemented on top of COMPSs. The \emph{dataClay} object store\cite{marti2017dataclay} is integrated with the framework and is able to provide active objects, providing data locality into the execution by leveraging the Python memory space of its backends.

Both COMPSs (\& dataClay) and Dask are perfect examples of task-based programming models that can take advantage of the \emph{SplIter}. They both support Python applications and have a similar set of features (chunked data structures, data locality, task-based full scheduling). Moreover, Dask offers a \emph{rechunk} mechanism that can be considered a competitor to the \emph{SplIter}.

Up until now we have discussed the state-of-the-art from the point of view of the programming model and the framework themselves. However, we must look into work related to the parallelization of loops and the scheduling of work. A foundational article on this field was published by Hummel \emph{et al.} with the Factoring\cite{hummel1992factoring} method. The work on Factoring has been extended with a focus on smarter and more complex scheduling mechanisms, e.g. with adaptative weighted factoring\cite{banicescu2001performance} or by adding also dynamic load balancing\cite{carino2008dynamic}. All this research shows the relevance of the granularity of tasks, an issue that we also aim to address. However, they tackle this issue for complex applications by discussing the method and its direct application into the application. Our focus is on a more generic \emph{SplIter} that coordinates with the task-based programming model and is aware of the data placement (for data locality purposes).


If we look at frameworks that expose lower level primitives we can find Charm++\cite{kale1993charm++}, a parallel system based on C++. This framework offers all the elemental primitives that we could expect, including iteration structures. However, Charm++ does not include neither scheduling mechanisms nor dependency management comparable to the ones available in commonly used task-based workflows (such as COMPSs or Dask).

\section{Environment architecture}

\label{sec:envarch}

We will describe and discuss two different software architectures: COMPSs~\& dataClay and Dask. Both are task-based frameworks, with similar design goals, that support Python and are suited for blocked data structures. Both frameworks will be used as baselines for the evaluation, and \emph{SplIter} is implemented and evaluated in both.

\subsection{COMPSs~\& dataClay}

\label{envarch:compssndataclay}

The COMPSs framework\cite{compss} provides a task-based programming model that can be used for the development of distributed applications. We have chosen it because it has a powerful and versatile scheduler as well as a runtime that is able to exploit the parallelism of applications.

Applications developed with COMPSs resemble sequential applications. This is by design: sequential programming is easy and a programming model focused on sequential development has  a welcoming learning curve. In this article we will be using PyCOMPSs\cite{tejedor2017pycompss} (the Python bindings) and the annotation of code is done as shown in Listing~\ref{lst:pycompsstask}.

\begin{lstlisting}[language=Python, caption=Use of \texttt{@task} decorator for defining PyCOMPSs tasks, label=lst:pycompsstask]
@task()
def increment(value):
  return value + 1
...
results = list()
for val in inputs:
  results.append(increment(val))
\end{lstlisting}

We can see that adding the \texttt{@task} decorator to a user function converts it to a task. The previous snippet shows a regular loop and a perfectly valid sequential application. However, by adding the \texttt{\@task} decorator and using the COMPSs framework, the information of the invocation reaches the scheduler. The framework then triggers an asynchronous operation (from the point of view of the application) resulting in a parallel loop. In addition to the asynchronous invocation of tasks, COMPSs will also manage the dependencies between tasks.

All the experiments that we will be showing in the evaluation make use of the \texttt{dislib}\cite{dislib} library. It is a distributed library implemented on top of PyCOMPSs. This library offers the blocked data structure we will use: the \emph{dislib array} --representing two-dimensional arrays, which are widely used in scientific applications and machine learning.

\subsubsection{dataClay}

We will be combining the COMPSs scheduler with dataClay~\cite{marti2017dataclay,dataclay-website}, a distributed object store that can be plugged into COMPSs. It offers additional features that have an impact on data locality, features that boost and have synergies with the \emph{SplIter}. Essentially, the implementation of \emph{SplIter} that we propose is implemented on top of dataClay; in the evaluation section on \ref{sec:eval} we will use the term COMPSs~\& dataClay to refer to this framework stack.

Internally, dataClay backends use the language native representation of the objects while the object is in memory. That is, dataClay Python backends will hold dataClay objects as Python objects within a Python interpreter. This approach ensures that the critical path of the applications will be able to leverage data locality, avoiding unnecessary data transfers of the dataset.

\subsection{Dask}

Dask\cite{daskweb} is a Python framework for parallel computing that combines a task-based scheduler with a set of data structures and algorithms for big data, analytics and HPC. The distributed library\cite{daskdistributed} offers further primitives appropriate for distributed computing and cluster-like environments.

There are several data structures provided by Dask (namely, Dask Array, Dask Bag, Dask DataFrame) but we will be focusing in the Dask Array (which resembles a \texttt{numpy.array}. The Dask Array has an application defined \emph{Chunk shape} value (which is the size of the blocks, called chunks, and sets the granularity). Their documentation already warns: ``Chunks should align with the computation that you want to do'', hinting the issues that an inadequate block size will harm performance.

Aside from the built-in data structures and their methods, the main primitive for defining and invoking tasks is the \texttt{delayed}. An example of how a function is defined as a task and how it can be invoked is shown in Listing~\ref{lst:daskdelayed}.

\begin{lstlisting}[language=Python, caption=Use of \texttt{@delayed} decorator for defining Dask tasks, label=lst:daskdelayed]
@dask.delayed
def increment(value):
  return value + 1
...
results = list()
for val in inputs:
  results.append(increment(val))
\end{lstlisting}

Calling \texttt{delayed} (directly or through a function that has been decorated) results in Dask generating nodes in a task graph. With this, Dask is able to ``run'' a graph and exploit parallelism and schedule tasks across the computing resources it has.

The \emph{Dask.distributed} library provides a more fined-grain primitive for task-scheduling: the \texttt{client.submit}. This mechanism is shown in Listing~\ref{lst:dasksubmit}. There are multiple methods similar to this one (e.g. the \texttt{client.map}). These methods, instead of working with a Dask graph, generate a \texttt{future} object for the computation they represent.

\begin{lstlisting}[language=Python, caption=Use of \texttt{client.submit} for invoking Dask tasks, label=lst:dasksubmit]
def increment(value):
  return value + 1
...
results = list()
for val in inputs:
  results.append(client.submit(increment, val, workers=w))
\end{lstlisting}

In addition to the task-scheduling aspect, Dask includes automatic mechanisms to leverage the data locality on its workers. The goal is to avoid data transfers and improve performance, and that is accomplished by caching Python objects in the Python memory space of the worker.

\subsubsection{rechunk}

\label{dask:rechunk}

The different Dask data structures interface offer some mechanisms to address the granularity, i.e. the size of its constituting blocks. One powerful operation is the \emph{rechunk} method of the Dask Arrays. This mechanism is a close competitor to the \emph{SplIter}: they both increase performance when the starting dataset is too fragmented, and they achieve that by effectively reducing the number of tasks.

The \emph{rechunk} creates a new data structure with a different chunk size. This can result in a lot of data transfers and in an increase of the memory footprint. The transfers occur because data is distributed, and the likelihood of locality between contiguous blocks decreases when the number of nodes increases. The memory footprint is affected by the duplication of data (required even in the best peer-to-peer strategy) for in-flight blocks and during the assembly of the new data structure.

We will compare and discuss the \emph{SplIter} performance to the Dask \emph{rechunk} approach in the evaluation.

\section{SplIter}

\label{sec:spliter}

\begin{figure}
    \centering
    \includegraphics[scale=0.55]{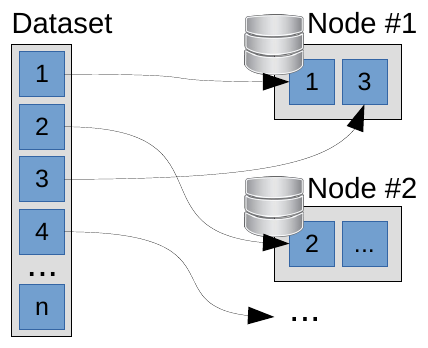}
    \caption{Diagram of the distribution of a dataset between different nodes}
    \label{fig:splitdiag-datasetdistribtion}
\end{figure}

When using a distributed execution environment there is the need of dividing the dataset into parts (i.e. blocks). This is required in order to distribute computation across nodes. Figure~\ref{fig:splitdiag-datasetdistribtion} shows a simplified diagram where we can see the meaning of the block distribution. The distribution may have additional requirements, or the data structure may carry more nuisances, but the main idea is the distribution of those blocks.

This block concept may be called with a variety of terms, and may be more or less transparent to the developer --for instance, the \emph{Resilient Distributed Dataset} or \emph{RDD}\cite{zaharia2012resilient} is a Spark data structure that transparently addresses the blocking needs of a distributed dataset and its distributed execution. The dislib array and the Dask array both use blocks and have a configurable and user-defined chunk size.

\begin{figure}
    \centering
    \includegraphics[scale=0.55]{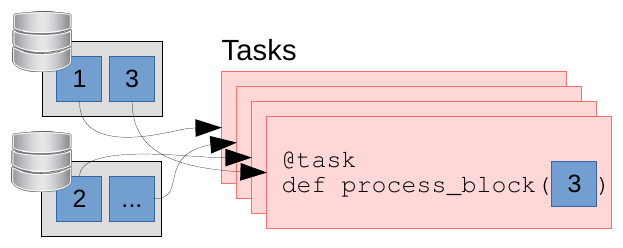}
    \caption{Diagram of a regular one-to-one execution of tasks to blocks (one task per block)}
    \label{fig:splitdiag-tasks}
\end{figure}

In task-based programming models, a usual scenario is to have a task for each block, as shown in Figure~\ref{fig:splitdiag-tasks}. Tasks may be more complex than that --e.g. by having multiple inputs. But, at a high level, we want to highlight the fact that the tasks will be using the blocks directly; this implies that the granularity of tasks depends on the block size. This is indeed what happens in both Dask and dislib algorithms.

Thus, applications need to define a block size, which is not trivially determined performance-wise. The level of fragmentation of the dataset will impact the scheduler overhead as well as the potential parallelism of the application. The optimal will depend on the implementation and the capabilities of the computing environment. Having a block per core is a good rule of thumb, but even that depends on the executing infrastructure (both on the number of nodes as well as on the CPU model in said nodes). Those parameters are not necessarily known during application development time. And even if everything is known beforehand, different stages of the application may have a different optimal block sizes (due to different numerical routines, elasticity of computing nodes, latency, etc.).

\begin{figure}
    \centering
    \includegraphics[scale=0.55]{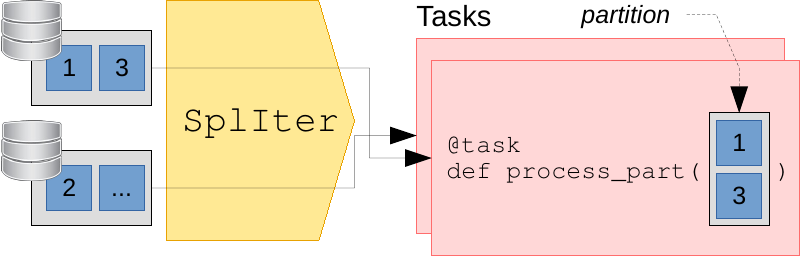}
    \caption{Diagram of \emph{SplIter} interactions and task invocation}
    \label{fig:splitdiag-splitandtasks}
\end{figure}

Our proposal consists on maintaining the blocks and enhance the iteration with the \emph{SplIter}, a software runtime mechanism that adapts both to the data placement of such blocks and to the computing capabilities of the environment. The \emph{SplIter} works as follows: gather all the blocks that are located in a single node and yield \textbf{partitions}. A diagram showing this procedure is shown in Figure~\ref{fig:splitdiag-splitandtasks}. Each partition is located in a single node, ensuring data locality, and the number of partitions (i.e. the number of tasks) is related to the computing capabilities of the environment.

In the COMPSs~\& dataClay stack, the \emph{SplIter} cooperates with dataClay in order to retrieve the location of objects and generate the partition objects (which will be also dataClay objects, linked to the same backend as the objects it represents). For Dask, we have implemented \emph{SplIter} through the use of the different \texttt{client} methods. The \emph{SplIter} implementation queries data location information to Dask and uses that information to both build partitions and to schedule the tasks according to the location of said partitions.

We will discuss the specific usage of \emph{SplIter} for each application in the following Section~\ref{sec:apps}. But before that, we want to illustrate the fundamental usage of the \emph{SplIter} and the codebase changes that it requires. We will be using here the Histogram as a sample application. The pseudo-code for the original implementation is shown in Listing~\ref{lst:original}. That implementation shows an embarrassingly parallel stage, called \texttt{partial\_\-histogram}, annotated with the \texttt{task} decorator; it also shows a merge stage --the \texttt{sum\_partials}, a direct addition-- annotated with the \texttt{task} as well as the \texttt{reduction} decorators. The pseudo-code used follows the general syntax of the COMPSs programming model (as shown in previous subsection~\ref{envarch:compssndataclay}), but the general idea behind this code is applied to Dask code.

\begin{lstlisting}[language=Python, caption=Sample of the original code --Histogram application, label=lst:original]
@task
def partial_histogram(block, ...):
    return np.histogramdd(block, ...)

@reduction    
@task
def sum_partials(partials):
    return np.sum(partials, axis=0)

# Main application
partials = list()
for block in experiment._blocks:
    partial = partial_histogram(block, ...)
    partials.append(partial)
result = sum_partials(partials)
\end{lstlisting}

The next Listing~\ref{lst:spliter} shows the changes required in order to make use of \emph{SplIter}. Note that the merge (\texttt{sum\_\-partials} function) remains unmodified; also the \texttt{partial\_\-histogram} remains the same --albeit it is not a task now but simply a function. What has been added is the \texttt{compute\_\-partition}, a task that processes all the blocks in a partition. The code in this inner loop (Listing~\ref{lst:spliter} lines 12-14) is the same as the code in the original loop (Listing~\ref{lst:original} lines 12-14). By design, adding the \emph{SplIter} results in an extra loop nesting as there is now an additional iteration per partitions (lines 19-21). The number of tasks (the number of \texttt{compute\_partition} invocations) is equal to the outer loop size (i.e. number of partitions).

\begin{lstlisting}[language=Python, caption=Minimal changes on sample code to include the usage of \emph{SplIter}, label=lst:spliter]
def partial_histogram(block, ...):
    return np.histogramdd(block, ...)

@reduction    
@task
def sum_partials(partials):
    return np.sum(partials, axis=0)

@task
def compute_partition(partition):
  part_results = list()
  for block in partition:
    partial = partial_histogram(block, ...)
    part_results.append(partial)
  return np.sum(part_results, axis=0)

# Main application
partials = list()
for partition in split(experiment):
  partial = compute_partition(partition)
  partials.append(partial)
result = sum_partials(partials)
\end{lstlisting}

As can be seen, using \emph{SplIter} does introduce some additional complexity, in the form of an additional loop. However, this pattern is simple enough and follows a clear semantics of the underlying abstractions: a loop for the partitions and a loop for the blocks that form the partition.

\subsection{Tracking collection order}

\label{spliter:getindexes}

Up until this point we have explained the foundation of the \emph{SplIter}. However, the mechanism that we have outlined so far loses the original ordering of the collection.

When the original collection ordering is relevant for the algorithm, we propose two different methods: \texttt{get\_indexes} and \texttt{get\_item\_indexes}. The first one, \texttt{get\_indexes}, returns the block index --so, in the example shown in Figure~\ref{fig:splitdiag-splitandtasks}, it would return \texttt{[1, 3]}. There are scenarios where the application requires the global element indexing; for those scenarios the \texttt{get\_\-item\_\-indexes} returns the individual item indexes.

By embedding index information into the partition, the algorithm can leverage the information of global block position and global item indexes. Some of the applications presented in this article will make use of those methods; we will mention the details for each application discussion --i.e. for the Cascade SVM (\ref{app:csvm}) and for the \emph{k}-Nearest Neighbors (\ref{app:knearestneighbors}).

\subsection{Comparison with rechunk}

There are several differences between the \emph{SplIter} and the \emph{rechunk} --what can be considered a direct competitor. The key difference is the fact that \emph{rechunk} generates a new array: once the \emph{rechunk} has finished, it has generated a new data structure with a different block size than the original. This contrasts with the \emph{SplIter}, which produces logical groups of blocks (the partitions). Using the \emph{rechunk} is simpler from the application point of view, as the result is a ``standalone'' array --same interface as the original dataset. But its simplicity comes with a penalty --when compared to the \emph{SplIter}-- in the form of additional transfers and data transformations.

Additionally, the \emph{SplIter} is able to leverage the data locality by producing partitions with worker-local blocks. This additional indirection (the partition) does not preserve the order, but this is addressed as described in the previous subsection~\ref{spliter:getindexes}.

\subsection{Implementation}

In this subsection we will review the general implementation steps that are required in order to recreate the \emph{SplIter} and we will discuss how these modifications have been introduced into COMPSs \& dataClay and Dask.

First of all, the \emph{SplIter} requires the data to be divided in blocks and distributed across nodes. The block subdivision is something that is already provided by the Dask Array and dislib Array data structures. The distribution of data across nodes is provided by Dask on one side and by dataClay in the other.

The \emph{SplIter} implementation will query the data location of the blocks and yield the \emph{partitions}. This procedure depends on the software stack and we will discuss them separately below. After the partitions have been assembled, we can proceed to the distributed execution step. In both frameworks (COMPSs and Dask) this will be done by invoking a task for each partition.

\subsubsection{COMPSs \& dataClay}

The first step was the dislib and dataClay integration. The \emph{SplIter} requires location information for distributed data, which is a feature provided by dataClay; the data blocks become dataClay objects and this requires some modifications in internal data manipulation functions in the dislib, but it does not affect the application interface of the array itself. The modifications in the dislib have been published in a public repository\cite{dislib-fork}.

Given a set of persistent objects --i.e., the blocks--, we can use dataClay to query the location of said objects. This is the beginning of the \emph{SplIter} implementation. With that information, the \emph{SplIter} implementation is able to create the partitions. The partitions are implemented as dataClay objects too --the partition is created in the same backend as its constituent blocks. This partition contains, fundamentally, the list of blocks (as node-local references, just as dataClay object references). Additionally, for the features explained in \ref{spliter:getindexes}, the partition contains the list of indexes (index for each block), and the list of item indexes (index for each element in each block). This information is populated during partition creation. The partition data structure can be found in the public application repository\cite{split-miniapps-github}.

Finally, a task that accepts the partition will be in charge of execution --task which will be invoked several times, as many as total partitions. Each task will iterate all the blocks in the partitions. This task is application-specific code. Note that \emph{SplIter} requires no modifications in the COMPSs framework.

The partition data model (including logic and index tracking) is about 50 lines of source code. The other modification required for implementing the \emph{SplIter} is the \texttt{spliter} function which amounts to less than 100 lines.

\subsubsection{Dask}

The Dask array is already a blocked data structure distributed across nodes --something that \emph{SplIter} requires. Moreover, the Dask API already provides a \texttt{who\_has} call which returns the location for a batch of objects in an efficient manner. As these features are already present in Dask, the \emph{SplIter} implementation has the appropriate data structure and the query mechanism.

With this features available, the \emph{SplIter} implementation can use them in order to generate the partitions. The partitions are, once again, built with the location query information. In this case, the partition references the blocks by the Dask identifier string. Each partition will contain the identifier of its objects. If indexing information is needed (as explained in \ref{spliter:getindexes}) it is generated along the partition.

When the execution is on the worker --i.e., when the task that processes a partition is being executed-- the actual object can be retrieved by using the identifier and the worker cache. More specifically, the worker cache is a dictionary of Python objects indexed by the Dask Future identifiers --the same identifiers that the partitions contain. This process (which is Dask specific) guarantees no data movements and also guarantees locality among tasks, partitions and blocks.

As this was done as a proof of concept, all required modifications were embedded into the application code (instead of modifying the Dask library). The codebase changes are available along the application code in the same application public repository\cite{split-miniapps-github}. That source code could be further documented and included into Dask upstream.

This implementation of the \emph{SplIter} is built with a minimal partition structure (a 15 lines long class definition) but requires accommodating certain Dask scheduling aspects, which adds 30-40 lines of source code. The \texttt{spliter} function is only 5 lines long, but once again, 30-40 additional lines of source code are required for managing the objects within the task.

\section{Applications}
\label{sec:apps}

This section will introduce the four different applications we will be using in our evaluation: Histogram, \emph{k}-means, Cascade SVM, and \emph{k}-Nearest Neighbors. Those specific applications have been chosen because they are commonly used algorithms and, on top of that, their data access patterns and implementation idioms can be seen in a whole lot of other different data analytics, machine learning and scientific applications.

All applications are implemented in Python. The main numerical library used in all of them is the NumPy library, which will be used either directly on the implementation or indirectly through higher-level abstractions. The source code for all the applications (as well as the results analysis) is public domain and is available on its public repository\cite{split-miniapps-github}. The datasets for all those applications are collections of \emph{n}-dimensional points. The data structure that will be used to hold them are either dislib native array objects or Dask native Dask Array structures.

They are both blocked data structures. Each block will represent a subset of points of the dataset and will, internally, be represented by a \texttt{numpy} two-dimensional array (with as many rows as points in the block and as many columns as dimensions per point). We will discuss the block size individually for each application in the evaluation section (Section~\ref{sec:eval}).

\subsection{Histogram}

\label{app:histogram}

We will start with a very iconic and fundamental application, an n-dimensional histogram kernel. We have chosen this application because of its simplicity and because it is a good representative of embarrassingly parallel applications that are very memory intensive --the Histogram operation is able to ingest huge chunks of data and process them in a short period of time. Examples of other kernels that share similar characteristics are filtering and aggregation operations.

For this application, we want to showcase the main advantages of the \emph{SplIter} and its effect on locality --which should be a key aspect on execution times due to the memory bandwidth bottleneck of the application.

The input dataset of this application (called \texttt{experiment} in the following code snippets) is a set of points. The embarrassingly parallel stage of the histogram evaluates a partial histogram for each block (using the \texttt{histogramdd} function of the NumPy library), and then the reduction stage merges those partial results into the final histogram result (with summation operations).

This application is memory intensive so data transfers play a key role on execution performance. We will discuss now the different implementations that we will be later evaluating in Subsection~\ref{eval:histogram}. First of all, the baseline code has been shown and discussed as the sample application in previous Section~\ref{sec:spliter} Listing~\ref{lst:original}. The code there shows: a) The main iteration across all the blocks --the main loop. b) The embarrassingly parallel task --the function \texttt{partial\_\-histogram} which processes a single block. c) The reduction task --called \texttt{sum\_\-partials}, which processes the partial results.

With this, we can move onto the \emph{SplIter} implementation, shown as the sample application in the previous Section (Listing~\ref{lst:spliter}). The summation operation done inside this \texttt{compute\_\-partition} (line 15) is the same operation done in \texttt{sum\_\-partials}; doing it within the \texttt{compute\_\-partition} task is done locally (guaranteeing that there are no data transfers for this first merge operation).

\subsection{\emph{k}-means}
\label{app:kmeans}

The next application is \emph{k}-means. It is a widely used clustering algorithm, sometimes used as a kernel within a bigger and more complex application. It is a good representative of machine learning algorithms, both in its semantic usage (clustering) as well as in its iterative implementation. We have chosen this application because it is based on a simple memory intensive microkernel at its core but it has certain complexity due to its reduction steps and overall iterative nature. This results in this application being a step above --in terms of complexity-- when compared to the previous Histogram, which shares certain memory-intensive aspects but it is much more fundamental.

For this application, we want to follow-up on the advantages of the \emph{SplIter} on memory intensive applications. The added complexity on the numerical procedures (the iterative nature of the algorithms, paired with a more costly reduction stage, compared to the previous application) will affect the execution times and the \emph{SplIter} contribution.

The input dataset of this application is a set of points. For each iteration, new centroids are evaluated by using the previous ones. This evaluation is done by calculating pairwise distances and aggregating points (which can be done in an embarrassingly parallel fashion) and evaluating the mean per each centroid (the merge stage).

This algorithm is called \emph{Lloyd's algorithm} (sometimes referred as \emph{standard algorithm}). Given its ubiquity and for the sake of brevity, we will refer to either the method, the algorithm, and the implementation, as simply \emph{k}-means. The reference implementation that we will be using is the one present in the \texttt{dislib} library.

We will now look into the implementation details and the required changes. First let's start with the original implementation, shown in Listing~\ref{lst:kmeans-original}. The main embarrassingly parallel task is the \texttt{\_partial\-\_sum} function. The reduction stage is encapsulated in the \texttt{\_recompute\-\_centers} call.

\begin{lstlisting}[language=Python, caption=Relevant lines of the original \emph{k}-means implementation (dislib codebase), label=lst:kmeans-original]
@task
def _partial_sum(row, old_centers):
    ...

class KMeans(BaseEstimator):
  def fit(self, x, y=None):
    ...
    while iterate:
      partials = list()
      for row_block in x:
        p = _partial_sum(row_block, old_centers)
        partials.append(p)
      self._recompute_centers(partials)
\end{lstlisting}

The modifications required for using the \emph{SplIter} are shown in Listing~\ref{lst:kmeans-spliter}. As discussed in Section~\ref{sec:spliter}, and similarly to the previous application, we can see the addition of a nested loop. In the outer loop we see, once again, a partial merge call (\texttt{\_merge} invocation). This function was already used by \texttt{\_recompute\-\_centers}, and now we can see its explicit invocation within the outer loop --an invocation with guaranteed locality.

\begin{lstlisting}[language=Python, caption=Relevant lines for using SplIter in the \emph{k}-means, label=lst:kmeans-spliter]
@task
def _partial_sum_partition(partition, centers):
    subresults = list()
    for block in partition:
        p = block.partial_sum(centers)
        subresults.append(p)
    return _merge(*subresults)

class KMeans(BaseEstimator):
  def fit(self, x, y=None):
    ...
    self.spl = spliter(x)
    while iterate:
      partials = list()
      for partition in self.spl:
        p = _partial_sum_partition(
            partition, old_centers)
        partials.append(p)
      self._recompute_centers(partials)
\end{lstlisting}

\subsection{Cascade SVM}

\label{app:csvm}

The next application that we will consider is a distributed SVM, or more precisely, the distributed training procedure of a support vector machine following the Cascade SVM algorithm. We have chosen this application because of its relevance as a data analytics algorithm as well as its importance in the machine learning ecosystem. The distributed implementation takes advantage of fundamental kernels (SVC) and globally it is a compute-bound algorithm. The evaluation and discussion for this application can be extrapolated to other applications with high computation requirements that start with an embarrassingly parallel stage and then have non trivial reduction procedures (e.g. mesh refinements algorithms, iterative optimization strategies, etc.).

For this application, we want to showcase the main advantages of the \emph{SplIter} when applied to a compute-bound application at its root. This is also an application where the item ordering is relevant for the result. The implementation that we will be using (the one in the \texttt{dislib} library) is based on the algorithm described by Graf \emph{et al.}\cite{graf2004parallel}.

The input dataset data structure is a set of points (called \texttt{x} later on in the code snippets). Along this array there is another one (called \texttt{y} in the pseudo-code), with the same cardinality, containing the \emph{labels} (or \emph{categories}) of those points --the \emph{SVM} is a supervised classifier. The Cascade SVM is an iterative algorithm, with each iteration starting by an embarrassingly parallel stage where a SVM is run in each block. The merge also consists on an SVM.

The actual implementation uses the \verb+sklearn+\cite{scikit-learn} C-Support Vector Classification (SVC). This can be considered the main microkernel of the application. As a result, the methods in this application have a high computation cost (with a low memory footprint).

Let us discuss the original source code of the \verb+CascadeSVM+ shown on Listing~\ref{lst:csvm-original}. In that snippet we can see the two arrays, one with points (parameter \texttt{x}) and the other with labels (parameter \texttt{y}).

\begin{lstlisting}[language=Python, caption=Relevant lines of the original CascadeSVM implementation (dislib codebase), label=lst:csvm-original]
@task
def _train(...)
    ...
    
class CascadeSVM(BaseEstimator):
  def _do_iteration(self, x, y, ...):
    for blockset in zip(...):
      x_data = blockset[0]._blocks
      y_data = blockset[1]._blocks
      _tmp = _train(x_data, y_data, ...)
\end{lstlisting}

The modifications introduced in the \emph{SplIter} implementation are shown Listing~\ref{lst:csvm-spliter}. The ordering within the collection matters, because the labels (parameter \texttt{y}) are linked to the points (parameter \texttt{x}). This is handled by using the \texttt{get\_indexes} mechanism, explained in subsection~\ref{spliter:getindexes}. The main task \texttt{\_train} is unmodified.

\begin{lstlisting}[language=Python, caption=Relevant modifications for the \emph{SplIter} implementation on the CascadeSVM application, label=lst:csvm-spliter]
@task
def _train(...)
  ...

class CascadeSVM(BaseEstimator):
  def _do_iteration(self, x, y, ...):
    for partition in spliter(x._blocks):
      spliter_indexes = partition.get_indexes()
      x_data = partition._chunks
      y_data = [y[idx]._blocks[0] 
                for idx in spliter_indexes]
      _tmp = _train(x_data, y_data, ...)
\end{lstlisting}

\subsection{\emph{k}-Nearest Neighbors}
\label{app:knearestneighbors}

The last application is the \emph{k}-Nearest Neighbors, an implementation of the non-parametric supervised learning method\cite{fix1951nonparametric}. The core of the algorithm is based on finding the \emph{k} elements that are nearest a certain point \emph{p}. The implementation has two distinct parts, named \emph{fit} and \emph{kneighbors} respectively, each with its own data. For each point of the \emph{kneighbors} dataset, its \emph{k}-Nearest Neighbors from the \emph{fit} dataset is returned --hence the algorithm name. Typical implementations (e.g. the one we are using from the \texttt{dislib}, which is directly using the \emph{sklearn} implementation) will generate tree lookup data structures during the \emph{fit} stage and those tree lookup data structures will be used to efficiently find neighbors during the second \emph{kneighbors} stage. 

It is an application where the result depends on the item ordering of the input, something that the \emph{SplIter} has to take into account. The support comes through the \texttt{get\_item\_indexes} primitive offered by our \emph{SplIter} proposal and discussed in subsection~\ref{spliter:getindexes}. 


For this application, we want to show the full potential of the \emph{SplIter} when applied to more complex data structures and algorithms. There will be more changes in the application --at least compared to the minimal ones shown in previous applications. However this will also unlock certain opportunities --as we will explain here and later on discuss during the evaluation, in subsection~\ref{eval:knearestneighbors}.

\begin{figure}
    \centering
    \includegraphics[scale=0.7]{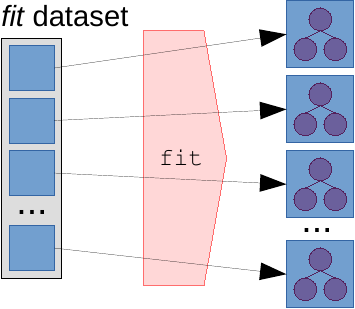}
    \caption{\emph{fit} stage of the \emph{k}-Nearest Neighbors application}
    \label{fig:knn-fit}
\end{figure}

As stated before, there are two input datasets for this application: \emph{fit} and \emph{kneighbors}. The \emph{fit} procedure takes the \emph{fit} dataset and generates the tree lookup data structures (one per block). This general procedure is illustrated in Figure~\ref{fig:knn-fit}. 

\begin{figure}
    \centering
    \includegraphics[scale=0.7]{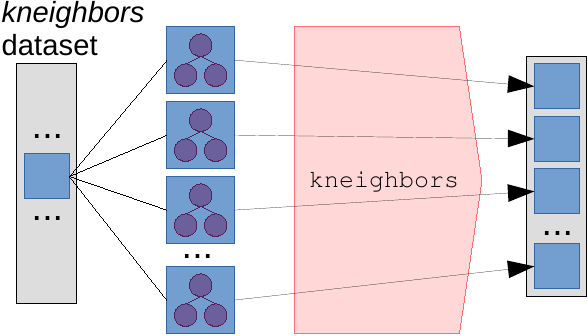}
    \caption{\emph{kneighbors} stage of the \emph{k}-Nearest Neighbors application (without depicting the final merge operation)}
    \label{fig:knn-kneighbors}
\end{figure}

The latter \emph{kneighbors} stage performs the lookup (for each input point on the \emph{kneighbors} dataset) against all the lookup trees. Figure~\ref{fig:knn-kneighbors} shows how a single block of the \emph{kneighbors} dataset is processed. The resulting number of tasks in this stage equals to the number of blocks in the \emph{kneighbors} dataset times the number of tree lookup structures we have. 

A final merge operation combines the partial results of the lookups into the result of the application.

Given the complexity increase for this application, we will look with more detail into the implementation and the distribution of tasks. Let's start with the original implementation, shown in Listing~\ref{lst:kneighbors-original}.

\begin{lstlisting}[language=Python, caption=Abridged relevant code of the original NearestNeighbors implementation (dislib codebase), label=lst:kneighbors-original]
class NearestNeighbors(BaseEstimator):
  def fit(self, x):
    for row_block in x:
      sknnstruct = _compute_fit(row_block)
      self._fit_data.append(sknnstruct)
      
  def kneighbors(self, y):
    indices = list()
    for q_row_b in y:
      queries = list()
      for sknnstruct in self._fit_data:
        q = _get_kneighbors(sknnstruct, q_row_b)
        queries.append(q)
      ind = _merge_kqueries(*queries)
      indices.append(ind)
    return indices
\end{lstlisting}

We have three distinct relevant tasks:

\begin{enumerate}
    \item \texttt{compute\_\-fit}, i.e. the tree generation. Internally, this operation is performed by calling \texttt{Nearest\-Neigh\-bors.\-fit} from the \texttt{sklearn} library\cite{scikit-learn}.
    \item \texttt{\_get\_kneighbors}, which evaluates a partial \emph{k}-nearest neighbors through the previous lookup trees.
    \item \texttt{\_merge\_kqueries}, which does the final merge stage by joining (sorting and picking) partial results.
\end{enumerate}

We show the \emph{SplIter} implementation in Listing~\ref{lst:kneighbors-spliter}. The general flow of the application is the same, with a difference in the \texttt{compute\_fit\_partition}, which now applies to a whole partition.

\begin{lstlisting}[language=Python, caption=Implementation of NearestNeighbors with \emph{SplIter}, label=lst:kneighbors-spliter]
class NearestNeighbors(BaseEstimator):
  def fit(self, x):
    for partition in spliter(x):
      nn = _compute_fit_partition(partition)
      self._fit_data.append(nn)
  
  def kneighbors(self, y):
    indices = list()
    for q_row_b in y:
      queries = list()
      for persistent_nn in self._fit_data:
        q = persistent_nn.get_kneighbors(q_row_b)
        queries.append(q)
      ind = _merge_kqueries(*queries)
    return indices
\end{lstlisting}

\begin{figure}
    \centering
    \includegraphics[scale=0.6]{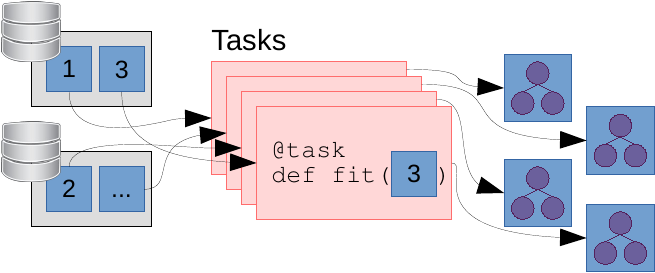}
    \caption{Diagram of the tree generation on the \emph{fit} (original implementation)}
    \label{fig:knn-diag_nosplit}
\end{figure}

\begin{figure}
    \centering
    \includegraphics[scale=0.6]{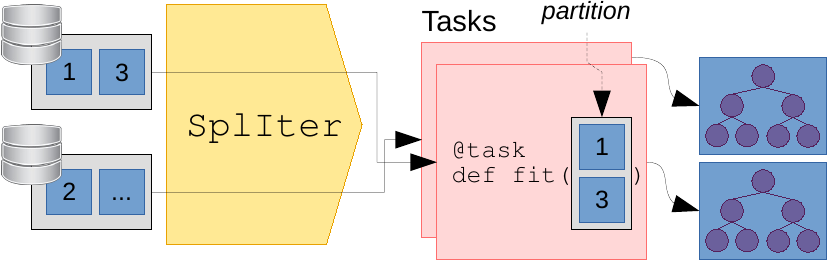}
    \caption{Diagram of the tree generation on the \emph{fit} (implementation with \emph{SplIter})}
    \label{fig:knn-diag_split}
\end{figure}

In the original \texttt{compute\_\-fit}, that task receives a block (conceptually, a set of points); on the other hand, \texttt{compute\_\-fit\_\-partition} processes a whole partition (which is a set of blocks, each block being a set of points, so at the end, a partition is also a set of points). This change seems minimal in the source code, but has an interesting implication: instead of generating a tree lookup data structure per input block (see Figure~\ref{fig:knn-diag_nosplit}), we are generating a single tree lookup data structure per partition --decoupling the number of intermediate data structures from the number of blocks in the input dataset. Figure~\ref{fig:knn-diag_split} shows the data structures resulting in the \emph{SplIter} version of the application. Using \emph{SplIter} in this fashion is done for two main reasons. First, it allows the implementation to exploit locality, as each tree is generated without requiring any serialization nor data transfer between nodes. Secondly, it generates more efficient lookup data structures, as having a single but bigger tree is more efficient on look-ups than having several smaller trees. Our evaluation in \ref{eval:knearestneighbors} will explore the impact of the tree sizes on number of tasks and total execution time. This efficiency increase due to different intermediate data structures shows an advantage that the \emph{SplIter} can bring us: by using partitions, the intermediate data structures can be combined and generated in a more sensible way.

Using \emph{SplIter} is not the only way to achieve bigger trees and the evaluation will consider the \emph{rechunk} approach as an alternative. The application developer could manage it with either hard-coding an optimal block size (which requires knowledge of the computing resources and lots of ugly hard-coding and platform-dependent code) or they could query manually the object store for the placement of objects and perform the partitions manually. The first option is an anti-pattern, and loses a lot of portability --negating the advantages of a task-based programming model. The second option is effectively reimplementing the \emph{SplIter} ad-hoc for each application, which is less efficient and yields no advantages over having it tightly integrated into the programming model --which is exactly what we propose.

\subsubsection{Parallel tree traversal}

Generally, Python numerical libraries manage the Global Interpreter Lock (GIL) in a smart way. This means that, even while Python is not able truly multithread the interpreter (because of this GIL) the numerical implementations can release the GIL and are typically multithreaded.

However, reading in parallel from a single data structure is more difficult. This happens in the \emph{kneighbors} stage of the application, during tree lookup operations. While technically it could be possible to implement the sklearn's KDTree implementation to be reentrant, this is not the current state.

To address that and be able to exploit parallelism, both implementations will include the snippet shown in Listing~\ref{lst:kneighbors-copytree} inside the \texttt{get\_kneighbors} task (during the \emph{kneighbors} stage). The solution consists of duplicating the iteration data structure (not the full tree) and that is achieved by performing a shallow copy of the \texttt{NearestNeighbors} data structure followed by a copy of its \texttt{\_tree} attribute (which is a KDTree instance).

\begin{lstlisting}[language=Python, caption=Shallow copy of sklearn data structure to allow parallel execution, label=lst:kneighbors-copytree]
    # nn is sklearn.neighbors.NearestNeighbors
    original = nn
    nn = copy(original)
    # nn._tree is a KDTree object
    nn._tree = copy(original._tree)
\end{lstlisting}

\section{Evaluation}
\label{sec:eval}

In this section we will be evaluating \emph{SplIter} performance for the different proposed applications. There will be two baseline executions, one per each framework; we will call the first one ``COMPSs~\& dataClay'' and the other one is called ``Dask''. These executions will be done without any mechanism to tackle granularity issues on the data --neither \emph{SplIter} nor \emph{rechunk}.

Besides these baselines, we will include both \emph{SplIter} executions (one on top of COMPSs~\& dataClay and the other one on top of Dask). An extra additional execution ``Dask~+ rechunk'' is included in the evaluation; this configuration will issue a \emph{rechunk} operation (see subsection~\ref{dask:rechunk}). This configuration is used as the main competitor of the \emph{SplIter}.

Applications have been executed several times (in the order of dozens to a hundred) in order to obtain meaningful measurements regarding their execution times. A winsorizing transformation is applied as to clean up outliers. The bar plots in this evaluation section will all include error bars that represent the percentile error showing the inter-quartile range. The percentile error is a nonparametric spread estimator that gives information regarding the spread or ``noisiness'' of execution times.

\subsection{Hardware}

\label{envarch:hw}

All the experiments are executed in the MareNostrum 4 HPC cluster~\cite{marenostrum}. The nodes in this cluster have the following technical specs:

\begin{itemize}
    \item 2$\times$Intel\textregistered{} Xeon\textregistered{} Platinum 8160L CPU @~2.10GHz
    \item 96GB of DRAM (12$\times$8GB 2667MHz DIMM)
    \item 100 Gb/s Intel Omni-Path (between computing nodes)
    \item 10 Gb Ethernet (storage and management)
\end{itemize}

Each node contains a total of 48 ($2 \times 24$) cores. The experiments in this article will be run in multiple nodes. The largest experiments will be done on a total of 16 computing nodes.

\subsection{Histogram}
\label{eval:histogram}

The first experiments that we will discuss are performed with the Histogram application, the most fundamental application that we will be discussing. It is a memory intensive application (lots of data and very fast execution). Because of this characteristics, a big dataset is used --but small enough to fit in RAM. To the memory footprint we have to also add the application execution itself as well as the framework. A size that could accommodate all our executions in all their different configurations was 880 million points (5 dimensions) per node. This results in a raw size of 33GB per node.

\subsubsection{Scalability for highly fragmented datasets}

We will start with a dataset divided into a large number of blocks. Our goal is to showcase the benefits of the \emph{SplIter} in an environment where locality and data transfers are critical --i.e. during a memory intensive application execution, such as the Histogram. The block size is chosen in order to attain 48 blocks per core.

\begin{figure}
    \centering
    \includegraphics[width=0.95\columnwidth]{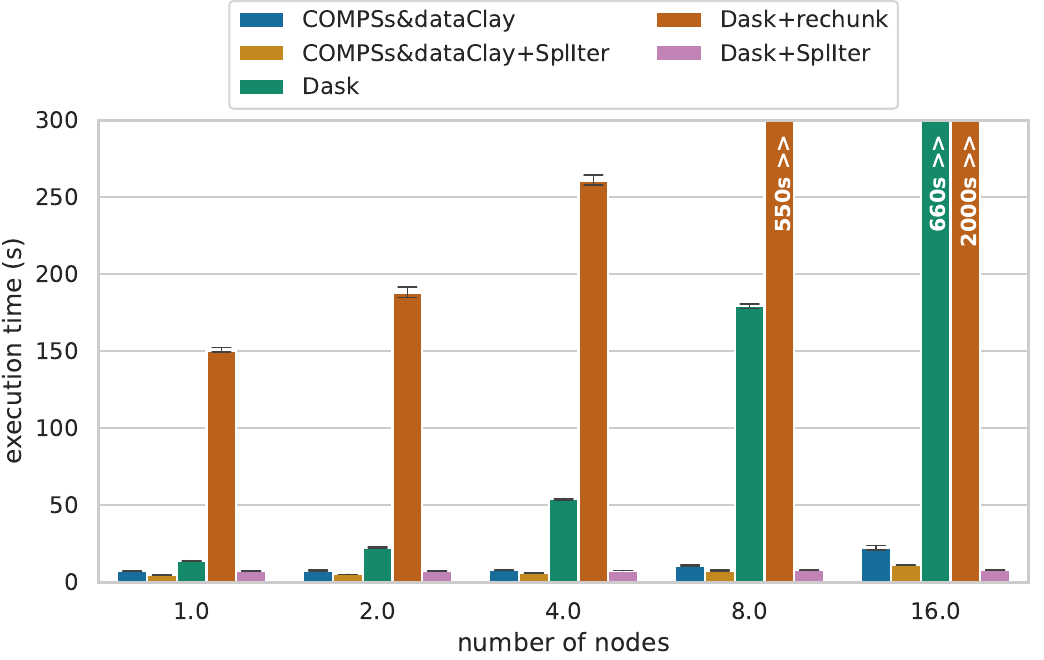}
    \caption{Histogram weak scaling from 1 node to 16 nodes, with 2304 blocks per node (48 per core)}
    \label{fig:histogram-weakscaling_smallblocks}
\end{figure}

The execution times are shown in Figure~\ref{fig:histogram-weakscaling_smallblocks}. 
The figure shows the good behavior of the \emph{SplIter} and its satisfactory scalability. Both \emph{SplIter} executions show similar execution times (meaning that the ``quantity of work'' done is similar, which means that both algorithm implementations are equivalent and the \emph{SplIter} approach is equally effective regardless the execution framework). These executions outperform their respective baselines for any number of nodes (and the difference grows with the number of nodes). This shows us that \emph{SplIter} is successfully reducing the scheduler and execution overhead, and the improvement increases with the number of blocks.

If we look into the scalability of the two baseline executions, we can observe that COMPSs~\& dataClay presents a good but not ideal scalability (i.e., the performance difference with \emph{SplIter} grows when the number of nodes and thus tasks grows). The performance degradation is due to the scheduling and invocation costs. The scalability of the Dask baseline is much worse. This suggests that the COMPSs scheduler is able to behave better under heavy pressure.

The competitor approach, Dask~+ rechunk, is a mechanism that reduces the number of tasks and improves the inadequate granularity issues. However, we can see in the figure that the execution times when using \emph{rechunk} are the worst. This happens in this kind of application due to the steep cost in terms of data transfers --a cost that shadows any improvement. The data transfers constitute a high expense in this application because the dataset size is large and the execution time is very low, meaning that data transfers become the main bottleneck instead of the computation or the scheduler overhead.

\subsubsection{Overhead for perfectly balanced datasets}

Now we move the evaluation to the scenario were the dataset is perfectly balanced; this means that there will be a single task per core. The total dataset size remains constant (880 millions points per node) so the change from the previous experiment is the number of blocks and the block size. Now the blocks are 18432 thousand points (48 times more than the previous experiment).

This experiment presents a worst-case scenario for the \emph{SplIter}, as a perfectly balanced dataset means that there is no room for improvement. Moreover, given that we are evaluating an application that has small absolute execution times, any penalty or additional noise will be conspicuous.

\begin{figure}
    \centering
    \includegraphics[width=0.95\columnwidth]{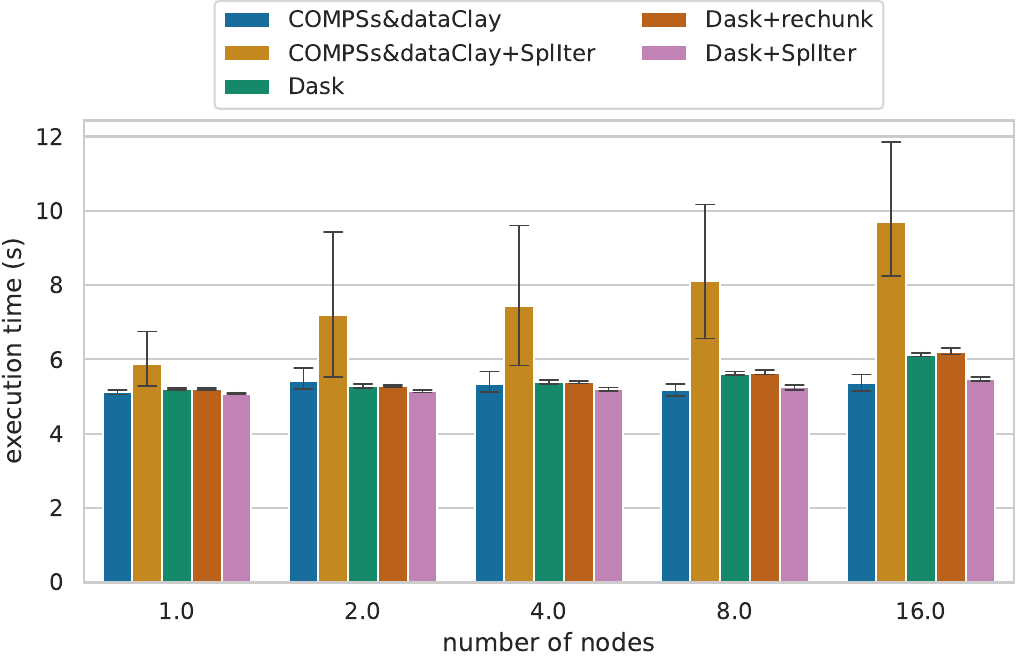}
    \caption{Histogram weak scaling from 1 node to 16 nodes, with 48 blocks per node (one per core)}
    \label{fig:histogram-weakscaling_bigblocks}
\end{figure}

The execution times for this experiment are shown in Figure~\ref{fig:histogram-weakscaling_bigblocks} where we can see a weak scaling from 1 to 16 nodes.

Excluding the COMPSs~\& dataClay~+ \emph{SplIter} execution, we see that all the other ones have close execution times between them. Both the baseline and the \emph{SplIter} are effectively performing the same Histogram operation unto the same data, and data is perfectly balanced.

In this specific scenario, the \emph{SplIter} implementation on Dask is able to exploit data locality a bit better; the \emph{SplIter} execution is able to exploit data locality during its kernel (just as its baseline) but the partitions are arranged divided in workers, which are able to exploit locality during the first reduction step (when partial results are merged together). Given that executions are so fast, being able to exploit this extra of data locality is able to give some additional benefit --a surprising result, given that this is designed to be a worst case scenario for \emph{SplIter}.

There is a slight difference among the two \emph{SplIter}, as we can see that the \emph{SplIter} implemented on top of Dask behaves much more consistently than the COMPSs~\& dataClay~+ \emph{SplIter} one. Of course, we have to take into account that this execution is very fast (just a few seconds long) but still, the overhead of the \emph{SplIter} on dataClay adds a fair amount of noise and an additional couple of seconds on average. Assembling the partition requires to retrieve location of blocks, and the overhead of that query will be different between dataClay and Dask. These results suggest that dataClay could be improved to have a faster query operation, and that would reduce the \emph{SplIter} overhead. Dask already offered a single operation to query the location of multiple objects, which --seeing the results-- is more efficient and consistent than the one we implemented on top of dataClay.

This worst-case scenario shows us that the overhead of the \emph{SplIter} --an overhead that depends on the number of blocks-- can be low. A higher number of blocks should result in a more laborious partition preparation, but the overhead can be minimal as demonstrated by the Dask execution.

\subsubsection{Sensitivity to fragmentation}

\begin{figure}
    \centering
    \includegraphics[width=0.95\columnwidth]{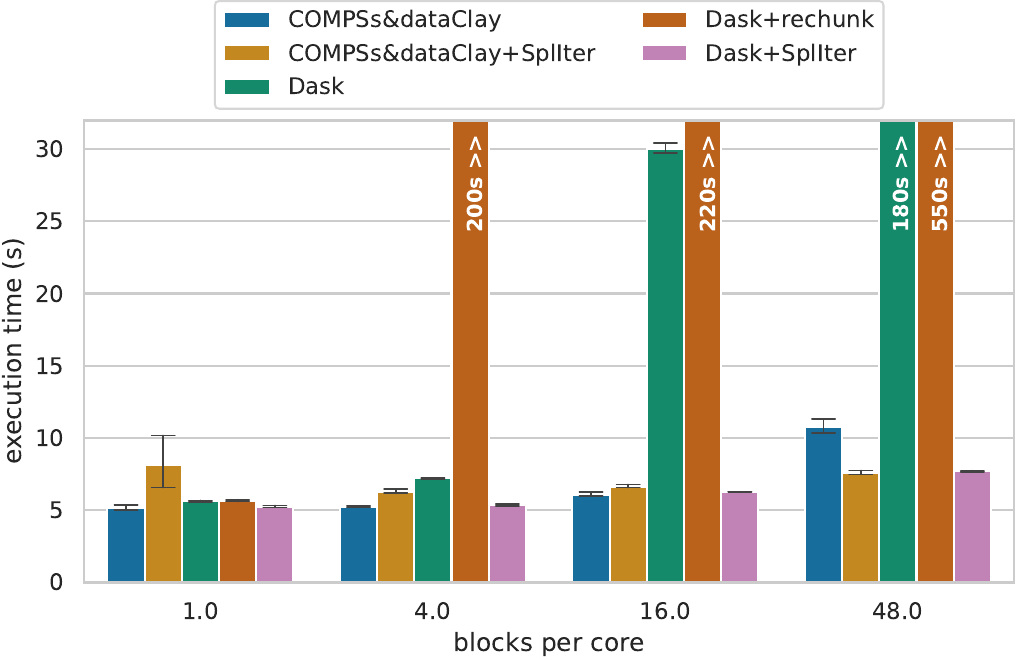}
    \caption{Histogram on 8 computing nodes. The X axis show variation on the total number of blocks per core. Block size changes in order for the total dataset size to remain constant.}
    \label{fig:histogram-blocksweep}
\end{figure}

After discussing a favorable scenario for \emph{SplIter} and an unfavorable one, we want to discuss what happens across this spectrum. This experiment will sweep the number of blocks from 1 per core (worst \emph{SplIter} scenario, which matches the previous experiment) to 48 blocks per core (best scenario, which matches the first experiment on this application). The experiment is evaluated in 8 nodes and the dataset size is constant and the same as in previous scenarios: 7 billion points in total. The execution times are shown in Figure~\ref{fig:histogram-blocksweep}.

The \emph{SplIter} behavior is quite flat, showing that it is a tool that desensitizes application execution performance from task granularity --the positive outcome that we expected. As we discussed for the worst-case scenario, the \emph{SplIter} implementation on top of dataClay has some room for improvement (the leftmost second bar, which stands out a little bit), but it shows good and steady performance everywhere else.

Once again we see that the baseline executions degrade a little bit (starting on the left being as fast as \emph{SplIter} and degrading progressively when the number of tasks increases. The COMPSs scheduler is much more robust against bad granularity scenarios (increasing from 5s to 10s) while the Dask scheduler suffers much more and increases its execution time from 5s to more than one hundred.

The rechunk is still a very expensive operation. When there is a single block per core, the rechunk operation is a no-op. However, as soon as the rechunk operation performs its duties (which happens at 4 blocks per core, when the rechunk is expected to redistribute data between workers) the huge amount of data that must be transferred between nodes results in hundreds of seconds of execution overhead.

We can see some slight inconsistencies between 1 and 4 blocks per core (the COMPSs~\& dataClay~+ SplIter implementation improves, while baseline executions are sitting very still). There is a performance variation that contributes to this: the special behavior of the microkernel (the \texttt{numpy.histogramdd} implementation that we are using in all scenarios) on big block sizes. The numerical implementation of the numpy \texttt{histogramdd} benefits from having smaller blocks: for instance, it is 20\% faster to perform 16 histograms of block sizes equal to 1152 thousand points instead of doing a single histogram of 18432 thousand points. This is taking into account reduction and independently of parallelism, just saturating all the cores and memory of a single socket. Discussing this application in this level of detail is outside the scope of this article, as it is related to the actual implementation on the \texttt{numpy} library and is also related to the intra-node architecture.

\subsubsection{Insights}

In this kind of memory intensive application, the \emph{SplIter} shows a good behavior, with big improvements on favorable scenarios and low-to-no overhead for unfavorable ones. This kind of application also renders the \emph{rechunk} approach impractical --using it results in an excess of data transfers. Also, the single-pass (as opposed to an iterative algorithm) negates the reuse potential of both \emph{rechunk} and \emph{SplIter} operations.

The \emph{SplIter} has shown that it is able to address the task granularity of the data while maintaining the data locality benefits; this is not achieved by the \emph{rechunk} for this kind of applications.

\subsection{\emph{k}-means}
\label{eval:kmeans}

The next application is \emph{k}-means, a memory bound iterative application. The dataset, when compared to the previous application, is a little smaller: 11GB of raw data per node (74 million points of 20-dimensional points per node). The \texttt{pairwise\_\-distances} function of the \texttt{sklearn} library --the main numerical function used in this application implementation-- is a memory hungry implementation which caused issues at bigger dataset sizes (specially for big blocks). We settled for this dataset sizes which ensures that all executions are consistent.

\subsubsection{Scalability for highly fragmented datasets}

Our goal for this experiment is to showcase the benefits of the \emph{SplIter} for another quite memory intensive application. This application has an iterative approach and more complex numerical procedures (when compared to the previous one). This first experiment is done with a highly fragmented dataset in which each block contains 64 thousand points and it is divided into 2304 blocks per node (which equals to 48 blocks per core).

The iterative approach should reduce the overall impact of both the \emph{rechunk} and the \emph{SplIter} approaches: the burden of those mechanisms are ``shared'' amongst all iterations, reducing its relative penalty. Moreover, the longer execution times will make the overheads less visible. The more complex reduction will have additional data locality benefits that the \emph{SplIter} can leverage during the reduction step.

We will be evaluating this scenario by performing a weak scaling, starting from 1 node up to 16. The experiment scales up to 36864 blocks for the 16-node execution.

\begin{figure}
    \centering
    \includegraphics[width=0.95\columnwidth]{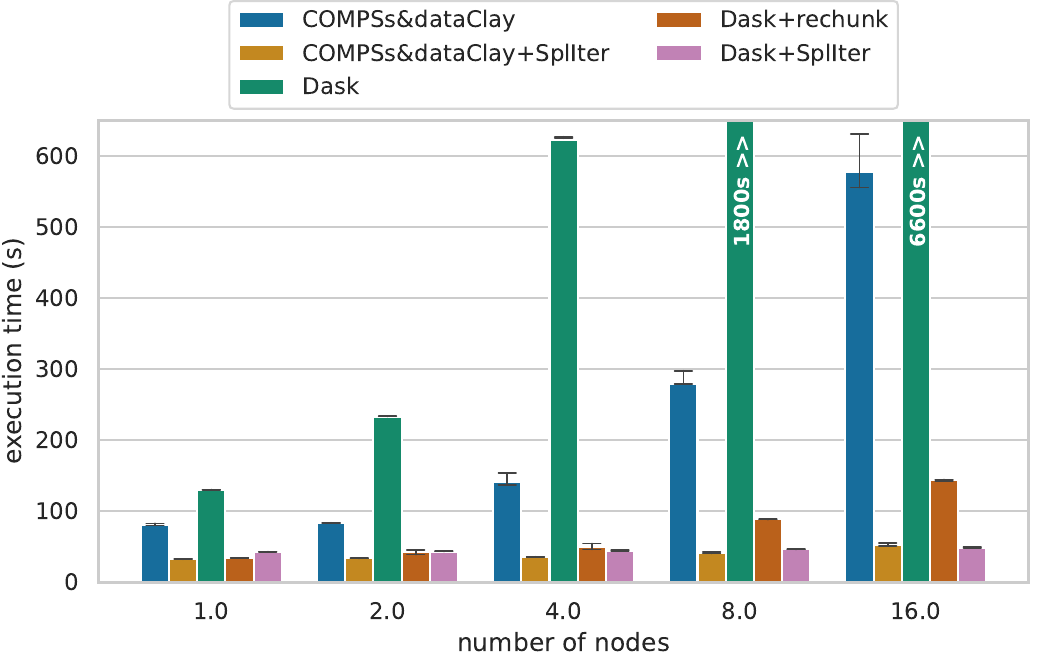}
    \caption{\emph{k}-means weak scaling from 1 node to 16 nodes, with 2304 blocks per node (48 per core)}
    \label{fig:kmeans-weakscaling_smallblocks}
\end{figure}

The execution times are shown in Figure~\ref{fig:kmeans-weakscaling_smallblocks}.
Once again we can observe that \emph{SplIter} executions have good scalability, showing very flat results and resulting in a very stable weak scaling behavior. Both implementations (COMPSs~\& dataClay~+ SplIter and Dask~+ SplIter) have results that are very close between them, showing that the amount of computation that they do is the same and the implementations are equivalent.

The baseline executions have a much worse scalability than the \emph{SplIter} counterparts. We already saw in the previous application (Histogram) that the scalability for highly fragmented datasets results in a degradation of performance due to the big amount of tasks. Given that we are evaluating an iterative application (10 loops in this evaluation), the overhead is amplified 10 times. The execution times of Dask end up being one order of magnitude slower, which shows that the COMPSs scheduler is more robust to being stressed with lots of tasks; however, even in that case the COMPSs~\& dataClay ends up being one order of magnitude slower than the \emph{SplIter} configuration.

An alternative to the \emph{SplIter} would be to use the \emph{rechunk}. This can be seen in the Dask~+ rechunk execution, which shows good results. Its scalability is not as good as the \emph{SplIter} and we can see a clear uptrend in the execution times when the number of nodes (and thus, the number of blocks) increases. This overhead is due to the rechunk cost, which requires to move data between workers (a cost that increases when the number of nodes increases). However, this cost is only payed once, not for every iteration (just as the \emph{SplIter} cost that is payed once).

\subsubsection{Overhead for perfectly balanced datasets}

The second experiment for this application will show what happens if the dataset is already perfectly balanced. Balanced means the same as in the previous experiment: the dataset is divided in such a way that there is one block per core. Once again, we want to focus on the overhead and general behaviour of the \emph{SplIter} on a worst-case scenario, one where there is no room for improvement for any \emph{enhanced iteration} mechanism.

\begin{figure}
    \centering
    \includegraphics[width=0.95\columnwidth]{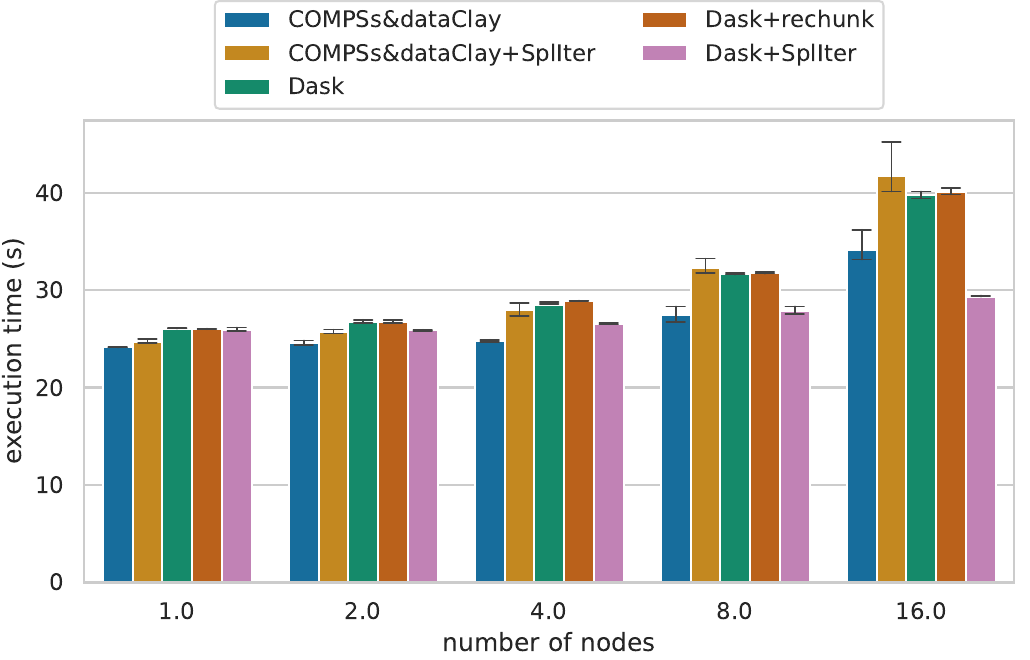}
    \caption{\emph{k}-means weak scaling from 1 node to 16 nodes, with 48 blocks per node (one per core)}
    \label{fig:kmeans-weakscaling_bigblocks}
\end{figure}

This scenario will also be conducted through a weak scaling. The dataset size will be identical to the previous scenario, but each block will be bigger: 3 million points per block ($48\times$ bigger than the previous scenario). Figure~\ref{fig:kmeans-weakscaling_bigblocks} shows the execution times for this scenario. 

We expected to see the pure overhead on the \emph{SplIter} executions. That is the case for COMPSs~\& dataClay~+ SplIter execution; the execution time grows with the number of nodes (i.e. blocks). However we see a much better behavior for the Dask~+ SplIter execution. As seen in the Histogram application, the overhead of \emph{SplIter} is lower in Dask. Moreover, the data locality achieved by the Dask workers during the reduction steps is higher than the one warranted by COMPSs.

In fact, the data locality achieved by the Dask~+ SplIter combination goes one step further. Because how the partitions are being generated (as detailed in Section~\ref{sec:spliter}), the first step of the reduction is able to exploit data locality too. This is not achieved by the baseline, and thus the Dask performance is slightly worse than the \emph{SplIter} one --specially for a high number of nodes as the reduction cost increases.

The \emph{rechunk} has no impact on execution times, as it is effectively a no-operation.

\subsubsection{Sensitivity to fragmentation}

We will follow with an experiment that should highlight the sensitivity of dataset fragmentation. The previous two experiments for this application have showcased a good case scenario and a worst case scenario for the \emph{SplIter}. Now we will explore this spectrum and find the tipping point where the benefits of the \emph{SplIter} are greater than its overhead.

\begin{figure}
    \centering
    \includegraphics[width=0.95\columnwidth]{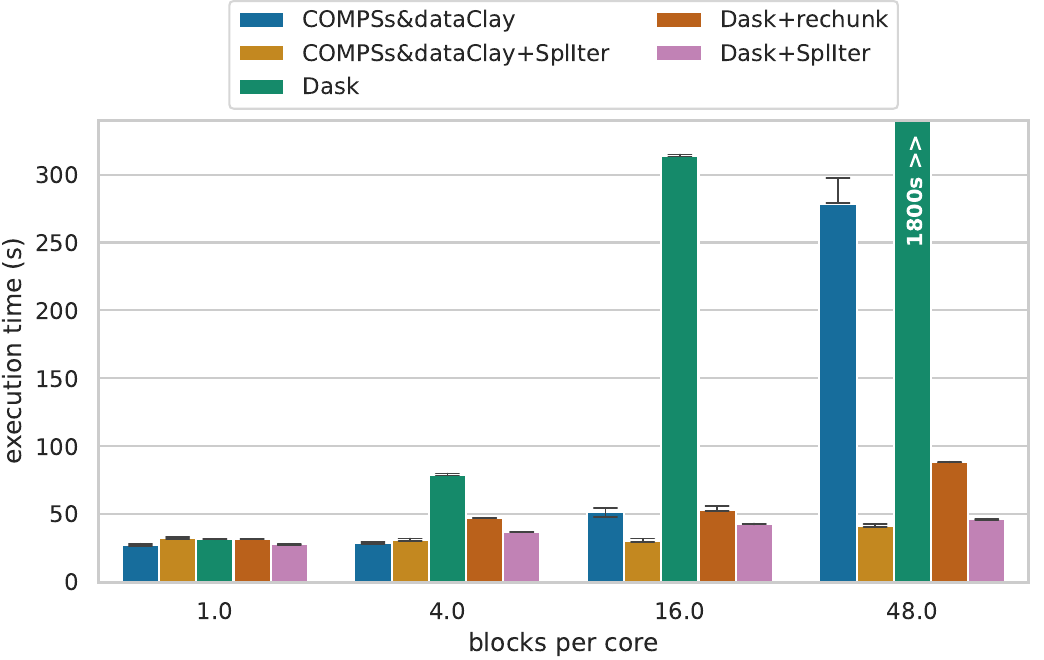}
    \caption{\emph{k}-means on 8 computing nodes. The X axis show variation on the total number of blocks per core. Block size changes in order for the total dataset size to remain constant.}
    \label{fig:kmeans-blocksweep}
\end{figure}

In order to do so we will once again fix the number of nodes to 8 and change the number of blocks per core. The number of blocks per core will go from 1 (worst case scenario for the \emph{SplIter}, as done in the previous experiment) up to 48 (which is the \emph{highly fragmented dataset} discussed in the first experiment for this application). The results can be seen in Figure~\ref{fig:kmeans-blocksweep}.

We have already discussed the leftmost and rightmost configurations (see the two previous experiments).

The \emph{SplIter} executions are very stable, almost flat, showing that this mechanism reduces sensitivity to fragmentation.

The baseline executions are sensitive to fragmentation and their execution performance decreases when the number of tasks increases --i.e. when the scheduler is under pressure and there are a lot of invocations of small tasks. COMPSs scheduler is able to ``resist'' a bit longer (it starts degrading significantly between 16 and 48 blocks per core) while Dask starts to increase at 4 blocks per core and quickly becomes orders of magnitude slower than any other execution.

The rechunk shows an overall good behavior. It does reduce sensitivity to the dataset fragmentation, but not as effectively as \emph{SplIter}. Its performance can be orders of magnitude better than the baseline but is still slower than the \emph{SplIter} execution.

\subsubsection{Insights}

When handling an iterative application such as the \emph{k}-means, including any mechanism to address the dataset fragmentation will result in substantial performance benefits. The overhead of \emph{SplIter} or \emph{rechunk} are diluted among the iterations and their benefits can be extremely large. Still, \emph{SplIter} outperforms the \emph{rechunk} because it is able to avoid data transfers (a cost that can grow for data intensive applications such as \emph{k}-means) as well as to maximize data locality.

\subsection{Cascade SVM}
\label{eval:csvm}

The next application is the Cascade SVM, a compute bound algorithm. The sizing of the dataset has been set in order to obtain reasonable execution times. Having a big dataset proved impractical for computation exploration as it required too many computing resources (i.e. for a statistically significant discussion). We have chosen a dataset size of 300 thousand points per computing node, which yields a good variety of execution times across all scenarios.

\subsubsection{Scalability for highly fragmented datasets}

As with previous applications, we will start with a highly fragmented dataset. Our goal is to showcase the benefits of the \emph{SplIter} for a compute-bound application. This means that the improvements due to locality will be lessened in comparison to previous applications. However, we do expect improvements due to the decrease in the number of tasks. By having less tasks we expect to increase responsiveness on the scheduler (it has less work to do) and a lower invocation overhead (due to the lower number of tasks). This is an ideal scenario for \emph{rechunk}, as the quantity of data that needs to be moved for a rechunking operation is small, while the improvements should be substantial due to the computation being the bottleneck.

\begin{figure}
    \centering
    \includegraphics[width=0.95\columnwidth]{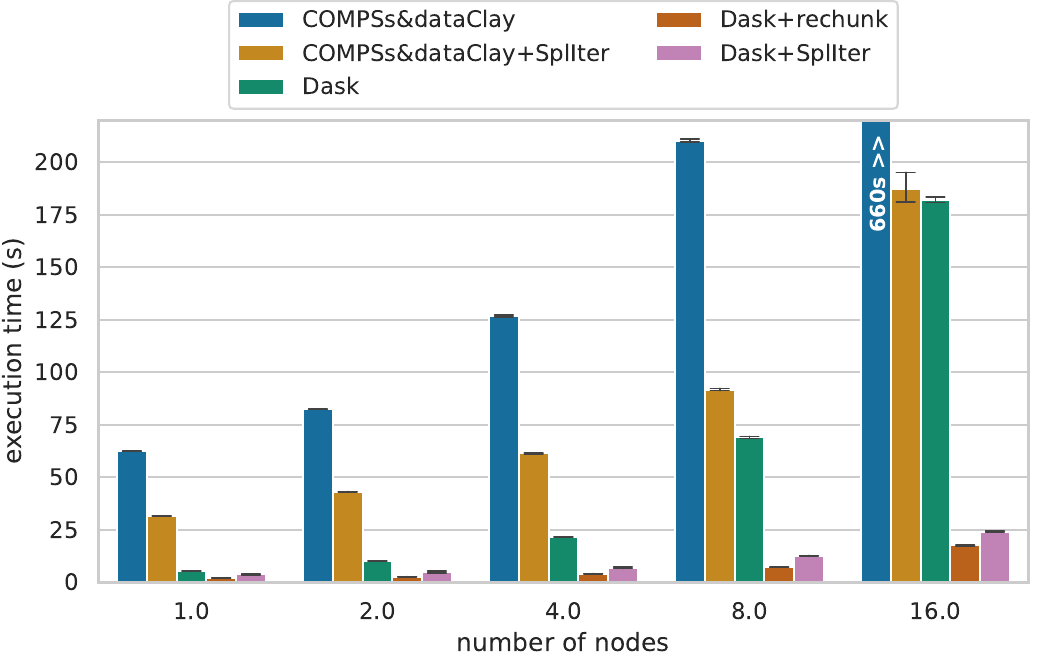}
    \caption{Cascade SVM weak scaling from 1 node to 16 nodes, with 384 blocks per node (8 per core)}
    \label{fig:csvm-weakscaling_smallblocks}
\end{figure}

The dataset is 300 thousand points per node. The evaluation will be performed by a weak scaling up to 16 nodes. The execution is performed with a block size of 128 points. This results in 8 blocks per core, or 384 blocks per node. The execution times for this scenario are shown in Figure~\ref{fig:csvm-weakscaling_smallblocks}.

Before analyzing the behavior of the \emph{SplIter} against the baseline and the rechunk, we should mention that there is a considerable difference between COMPSs and Dask. This difference is related to the baseline implementations and not to the \emph{SplIter} or to the data granularity. The current execution (highly fragmented dataset) has several factors that make this comparison complex, so we will discuss this in the next configuration with a perfectly balanced dataset.

The \emph{SplIter} executions outperform their respective baselines substantially. The improvement introduced by the \emph{SplIter} is apparent, once again, for a highly fragmented dataset.

For this application, we see how the \emph{rechunk} outperforms the \emph{SplIter}. As discussed in \ref{dask:rechunk}, the \emph{rechunk} materializes a new array, while the \emph{SplIter} does not. In previous applications, the data transfer was bigger and resulted in faster execution times for \emph{SplIter}. In this application, however, the data transfers are small enough to be almost invisible, while having the array materialized results in faster execution times.

\subsubsection{Overhead for perfectly balanced datasets}

We will now show an evaluation for a perfectly balanced dataset --with as many blocks as there are cores. In that scenario, once again, there should be no improvement for the \emph{SplIter} or the \emph{rechunk} --there is already a one-to-one relationship between computing resources (cores) and tasks (blocks), so there is nothing to improve in that regard. This evaluation will help us evaluate and characterize the overhead in extreme cases where no potential benefit exists from the point of view of the enhanced iteration mechanism.

This experiment will maintain the same dataset size (300 thousand points). When comparing to the previous scenario, the block size is increased from from 128 to 1024, and the number of blocks per core is reduced from 8 (the previous ratio) to 1 (i.e. perfectly balanced).

\begin{figure}
    \centering
    \includegraphics[width=0.95\columnwidth]{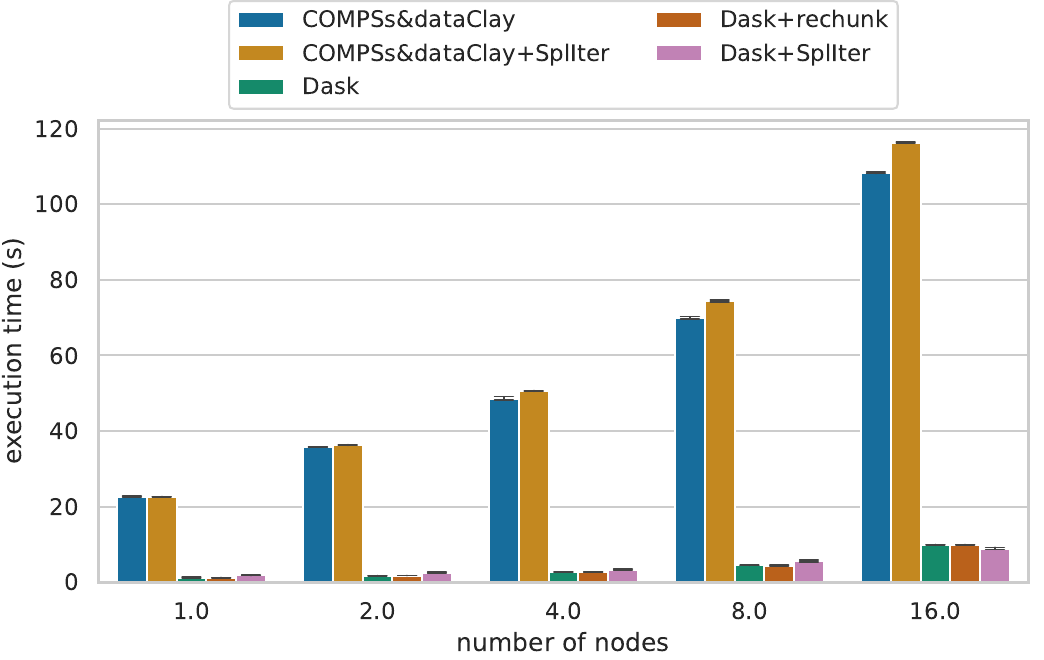}
    \caption{Cascade SVM weak scaling from 1 node to 16 nodes, with  48 blocks per node (one per core)}
    \label{fig:csvm-weakscaling_bigblocks}
\end{figure}

Figure~\ref{fig:csvm-weakscaling_bigblocks} shows the execution times for this scenario.

Before discussing the \emph{SplIter}, we can clearly see the different performance between COMPSs and Dask. Up until now, prior application performances have been very close between COMPSs and Dask, as we have made an effort to compare equivalent implementations. In this case, however, some details of the implementation have resulted in a disparity of performance; both algorithms are effectively doing the same operations (we used the \texttt{dislib} implementation as the reference, and implemented that in Dask) but the general data management of intermediate results and the internal serializations are causing this difference. To address that we could try to improve the \texttt{dislib} implementation --which may prove difficult and is outside the scope of this article. Even if the current results hinder simultaneous comparisons across COMPSs and Dask, we can discuss the \emph{SplIter} impact separately for both frameworks.

Both \emph{SplIter} executions follow closely their respective baselines. The Dask~+ rechunk is almost exactly the same as Dask. The overhead on the COMPSs~\& dataClay executions is more visible, and we can observe how the overhead depends on the number of blocks --i.e. the difference between the baseline and the \emph{SplIter} increases when the number of nodes increases. This overhead is smaller in Dask (a phenomenon that we have already seen in previous applications) which suggests that the mechanism to query location and build partitions is more efficient in Dask than in dataClay.

The \emph{rechunk} is effectively a no-operation and its performance is identical to the regular Dask.

\subsubsection{Sensitivity to fragmentation}

We have shown a relatively beneficial scenario as well as a worst-case scenario from the point of view of the \emph{SplIter} and \emph{rechunk} mechanisms. In the following experiment we will showcase what happens across this spectrum, and specifically we will discuss how sensitive are the enhanced iteration techniques to the quantity of blocks and fragmentation of the dataset.

This experiment will fix the dataset size, fix the number of computation nodes, and vary the number of blocks. More tasks results in a more saturated scheduler and more runtime overheads --which translates to more opportunities for mechanisms that addresses granularity such as the \emph{SplIter} or the \emph{rechunk}.

\begin{figure}
    \centering
    \includegraphics[width=0.95\columnwidth]{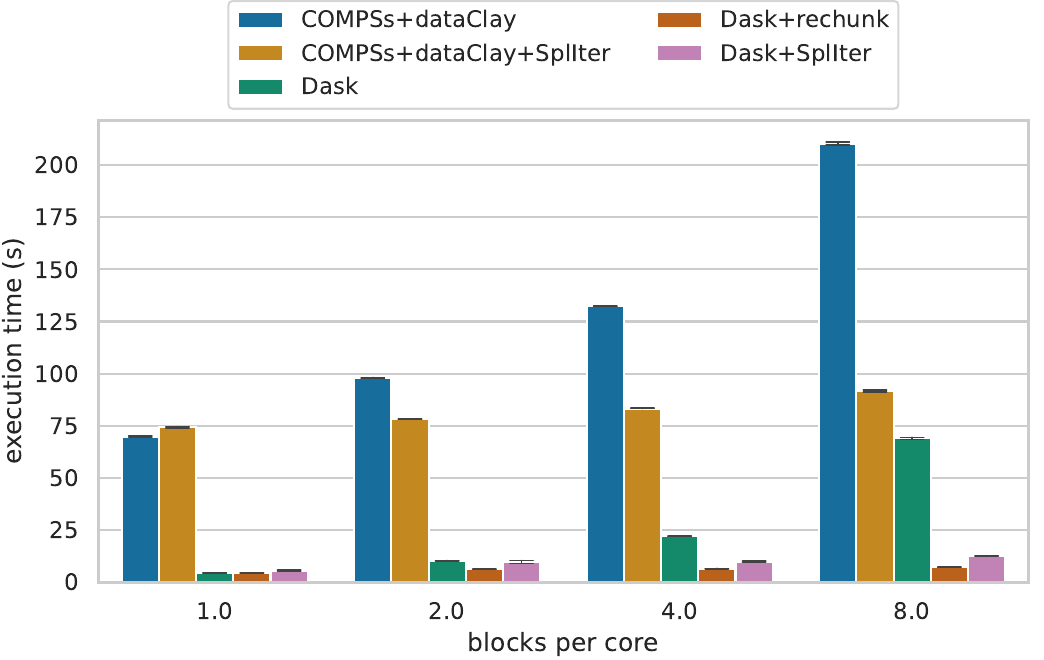}
    \caption{Cascade SVM on 8 computing nodes. The X axis show variation on the total number of blocks per core. Block size changes in order for the total dataset size to remain constant.}
    \label{fig:csvm-blocksweep}
\end{figure}

The execution is spread across 8 computing nodes. The execution times can be seen in Figure~\ref{fig:csvm-blocksweep}. The leftmost group of bars show the execution times when the dataset is perfectly balanced (i.e., previous experiment); the rightmost group of bars show the other end of the spectrum (i.e., the second to last experiment, which corresponds to 8 blocks per core).

The \emph{SplIter} executions are almost flat, showing that the \emph{SplIter} is reducing the sensitivity to the task granularity. The execution times for 1 block per core (the worst case for the \emph{SplIter}) is almost the same as the execution time for 8 blocks per core.

On the other hand, we see that both COMPSs~\& dataClay and Dask are quite sensitive to fragmented datasets. Their execution times are equal to \emph{SplIter} for 1 block per core and grow to more than double when the fragmentation increases to 8 blocks per core. All this overhead is due to the stress on the scheduler and the high number of task invocation, which become the bottleneck of execution.

This application is perfect for \emph{rechunk}. As soon as there is more than one block per core, \emph{rechunk} becomes the fastest configuration. The relatively small dataset results in a fast \emph{rechunk} operation --meaning that the overhead of performing a rechunk is almost irrelevant-- while having a materialized collection results in faster execution times when compared to the unmaterialized collection --the partitions yielded by the \emph{SplIter}.

\subsubsection{Insights}

This is the first compute bound application that we have discussed and, once again, we have seen that there is the need to address dataset fragmentation. In this application, \emph{rechunk} yields better performance than \emph{SplIter}. The two previous applications where memory intensive and \emph{SplIter} outperformed \emph{rechunk}; this application, a CPU intensive one, shows that \emph{rechunk} can beat the \emph{SplIter} in certain scenarios. Even in these scenarios much more favorable to the \emph{rechunk}, the difference between the two mechanisms is small.

\subsection{\emph{k}-Nearest Neighbors}
\label{eval:knearestneighbors}

This application has two different datasets, both used as inputs: the \emph{fit} dataset and the \emph{kneighbors} dataset. In our executions, each block contains 500 thousand three-dimensional points --a manageable size, but big enough to result in over one second long single tasks. Using three-dimensional points is a natural choice for point clouds, but the implementation and our conclusions are generic and applicable also for higher dimensional points (that may be used in classification or regression).

Given that this application presents a more convoluted algorithm consisting of two distinct parts, we will start by discussing the behavior of the two main microkernels that are used in the application: the \emph{fit} procedure and the \emph{kneighbors} procedure. After this first experiment, we will discuss the full software stack with two more experiments; the first one will be a scalability with a fixed ratio between the two datasets; the latter will perform an analysis when the \emph{fit} dataset is scaled.

\subsubsection{Kernel characterization}

When discussing this application in~\ref{app:knearestneighbors} we discussed the impact that the size of the tree data structures will have onto the algorithm. There are a lot of nuisances, but the general intuition is that using big trees will result in less tasks, more efficiency and less overall execution times. In this first experiment we will show the impact that the tree size (which is directly related to the \emph{fit} dataset block size) has onto the two main kernels of the application (i.e. the \emph{fit} stage and the \emph{kneighbors} stage. Each stage purpose and characteristics have been discussed in \ref{app:knearestneighbors} and more specifically outlined in Figure~\ref{fig:knn-fit} and Figure~\ref{fig:knn-kneighbors}.

\begin{figure}
    \centering
    \includegraphics[width=0.85\columnwidth]{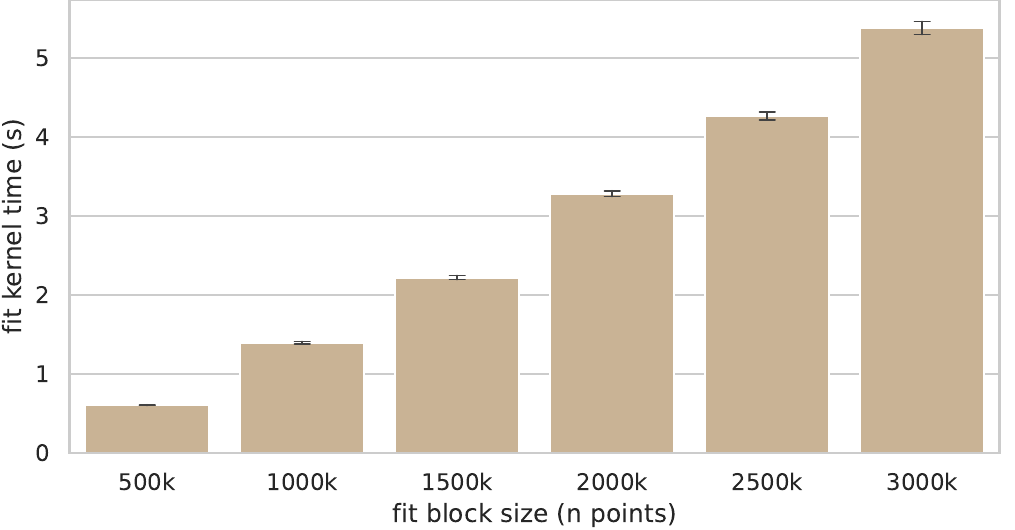}
    \caption{Kernel processing time for the \emph{fit} procedure of the \emph{k}-Nearest Neighbors application while increasing the \emph{fit} block size}
    \label{fig:kneighbors-kernel-fit}
\end{figure}

\begin{figure}
    \centering
    \includegraphics[width=0.85\columnwidth]{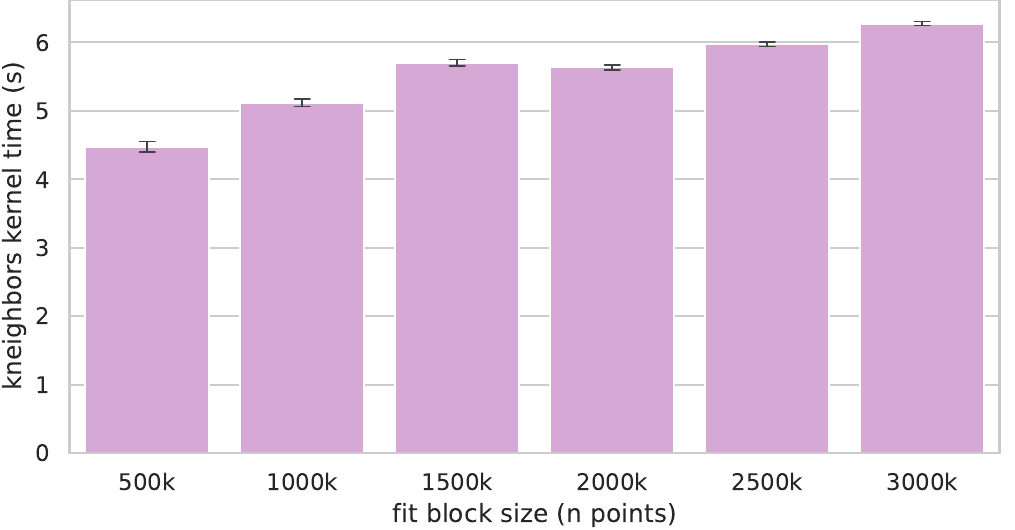}
    \caption{Kernel processing time for the \emph{kneighbors} procedure of the \emph{k}-Nearest Neighbors application while increasing the \emph{fit} block size}
    \label{fig:kneighbors-kernel-kneighbors}
\end{figure}

The times that we will be evaluating correspond to the numerical execution time for the payload of a single task (single block). In our scenario, this means executing the code on \emph{sklearn} library without parallelism, data transfers nor framework overhead. As both stages are sensitive to the block size of the \emph{fit} dataset, we will show the execution times when varying the \emph{fit} block size. Figure~\ref{fig:kneighbors-kernel-fit} shows the execution times of the \emph{fit} kernel and Figure~\ref{fig:kneighbors-kernel-kneighbors} shows the execution times for the \emph{kneighbors} kernel.

For the \emph{fit} kernel (Figure~\ref{fig:kneighbors-kernel-fit}) we can see how, when the input block increases, so does the processing time --in an almost linearly fashion. We are seeing here the cost of building the tree data structures and the cost is as expected.

The behavior of the \emph{kneighbors} kernel (Figure~\ref{fig:kneighbors-kernel-kneighbors}) is much flatter. We already expected this (as previously discussed in~\ref{app:knearestneighbors}) because tree lookup operations do not increase linearly.

Table~\ref{tbl:knnprojection} explores microkernel execution time for four different block sizes. The first column shows the block size of the \emph{fit} dataset (ranging from 3000k points for the biggest one down to 500k for the smallest). The second column shows the number of blocks in the \emph{fit} dataset (equivalent to the number of tasks). This is a parametric value, which will depend on the size of the \emph{fit} dataset. Decreasing the block size results in a bigger number of blocks; e.g. given a starting dataset of $n$ blocks with a block size of 3000k points, that same dataset will yield $3n$ blocks when the block size equals 1000k points. The \emph{fit} total time (meaning the sum of all kernel execution times, sequentially) also depends on $n$ and varies depending on the block size. The \emph{kneighbors} stage has its own dataset, which is why the last column on the table shows the execution time \emph{per block}; the time shown in the table is the sum of all the \emph{kneighbors} kernel executions (sequentially) for a single \emph{kneighbors} dataset block. To get the \emph{kneighbors} total time we would need to multiply that value per the number of blocks in the \emph{kneighbors} dataset.

\begin{table}[]
\center
\begin{tabular}{rrrr}
\multicolumn{1}{l}{\textbf{Fit}} &
\multicolumn{1}{l}{\textbf{\#\emph{fit} tasks}} &
\multicolumn{1}{l}{\textbf{\emph{fit}}} &
\multicolumn{1}{l}{\textbf{\emph{kneighbors}}} \\

\multicolumn{1}{l}{\textbf{block size}} &
\multicolumn{1}{l}{\textbf{(= \#blocks)}} &
\multicolumn{1}{l}{\textbf{total time}} &
\multicolumn{1}{l}{\textbf{time per block}} \\
\hline \hline
3000k & $n$ & $n\cdot$ 5.38s & $n\cdot$ 6.28s \\ 
1500k & $2n$ & $n\cdot$ 4.44s & $n\cdot$ 11.42s \\ 
1000k & $3n$ & $n\cdot$ 4.17s & $n\cdot$ 15.35s \\ 
500k & $6n$ & $n\cdot$ 3.66s & $n\cdot$ 26.83s \\ 
\end{tabular}
\caption{Combined execution times for \emph{k}-Nearest Neighbors microkernels.}
\label{tbl:knnprojection}
\end{table}

The table shows how the \emph{kneighbors} is the longest stage and also the most sensitive to the block size (i.e. tree size). It is natural to consider beneficial configurations for it, and those are found when the \emph{fit} block size are big --i.e. when the tree data structures are big. Moreover, the tree data structures generated during the \emph{fit} may be used by a lot of \emph{kneighbors} blocks, and those trees may be used repeatedly during execution (e.g. for iterative algorithms); these characteristics dilute the cost of the \emph{fit} stage when compared to the \emph{kneighbors} stage.

With that in mind we will set up both \emph{SplIter} and \emph{rechunk} to build a single tree lookup data structure per location. This prior analysis corroborates the intuition that bigger \emph{fit} block sizes increase the performance of the microkernel execution time. The last experiment for this application (see~\ref{kneighbors:speedup}) will revisit this same discussion on the \emph{kneighbors} execution time, evaluated in the complete stack environment (with COMPSs, with multiple nodes, with parallelism, and with full datasets).

\subsubsection{Scalability}

In this experiment we will include the software stack and evaluate the scalability and parallelism of the \emph{SplIter} and \emph{rechunk} mechanisms for a high quantity of blocks and tasks. Our goal is to evaluate our proposal in the context of a complex application, application which also contains non-trivial data reuse. We already explained certain implementation details required in order to change the tree lookup structures size.

\begin{figure}
    \centering
    \includegraphics[width=0.95\columnwidth]{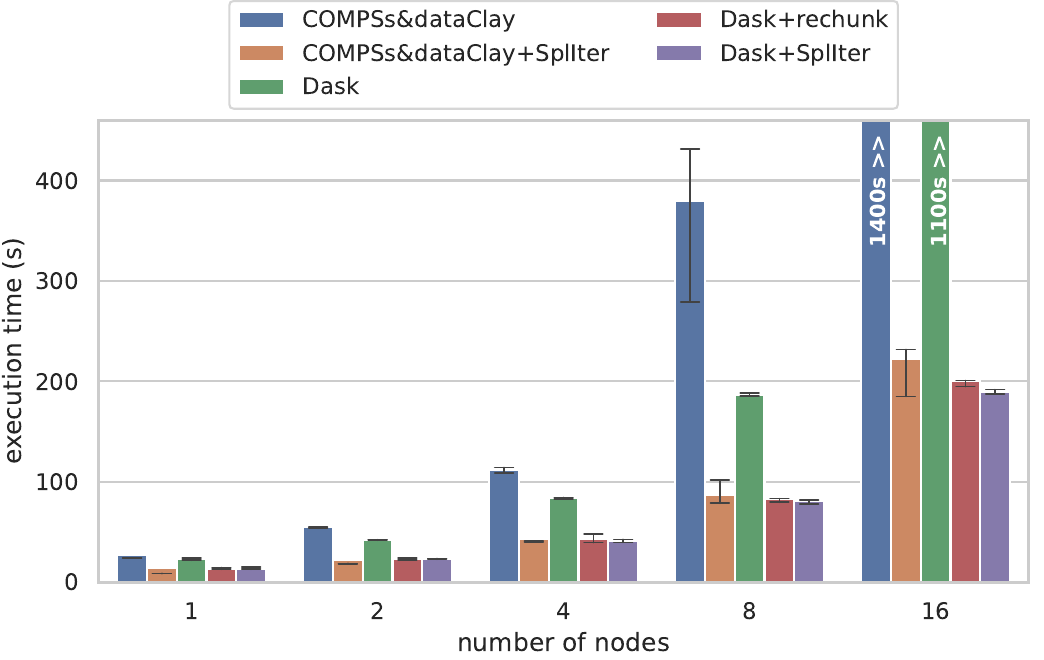}
    \caption{\emph{k}-Nearest Neighbors scaling from from 1 node to 16 nodes}
    \label{fig:kneighbors-weakscaling}
\end{figure}

The general scalability scenario will be evaluated by using the \emph{fit} dataset at 6 blocks per node, and 24 blocks per node for the \emph{kneighbors} dataset. Consequently, the \emph{kneighbors} stage will consist of a total of $6 \times 24 \times n_{nodes}$ tasks for baseline executions. Regarding the other executions, we will follow the conclusions reached in the previous experiment and use a single tree per location, i.e. per backend/worker. In our NUMA architecture computing environment this means to have a single tree per socket. This results in a total of $2 \times 24 \times n_{nodes}$ tasks for both \emph{SplIter} and \emph{rechunk} executions. The result of this first experiment are shown in Figure~\ref{fig:kneighbors-weakscaling}.

The \emph{SplIter} executions are very close among them, showing that they perform the same computations and the \emph{SplIter} performance is similar in both frameworks. The \emph{k}-Nearest Neighbor cost increases when the problem size increases (i.e. when the number of nodes increases), which results in an increase in execution times in the \emph{SplIter} executions.

The baseline executions increase faster, showing a quicker degradation of performance. In previous applications we have seen that the COMPSs framework is more robust against scheduler pressure. However, the number of tasks is not as big in this application, and the data locality is more relevant --a feature that Dask is able to leverage more efficiently than COMPSs, due how Dask manages in-memory Python objects. This explains why the COMPSs~\& dataClay baseline is slower than Dask.

The \emph{rechunk} execution is virtually identical to the \emph{SplIter}; this was expected given that the final tree data structures should be equivalent (meaning equally large) in both scenarios.

\subsubsection{Speedup for \emph{fit} dataset scaling}
\label{kneighbors:speedup}

In this experiment we will be scaling the training dataset (the first one, the one that is fed to the \emph{fit} stage). The number of nodes is fixed to 8 and the size of the second dataset is also fixed to 24 blocks per worker, which is a total of 192 blocks.

Our goal is to observe the impact of both the \emph{SplIter} and the \emph{rechunk}, and how well they scale when the data structures generated in the \emph{fit} stage grow. In these experiments, the non-baseline executions will generate a tree lookup data structure per location (i.e. one per backend/worker, or one per socket, which equals to a total of 16).

\begin{figure}
    \centering
    \includegraphics[width=0.95\columnwidth]{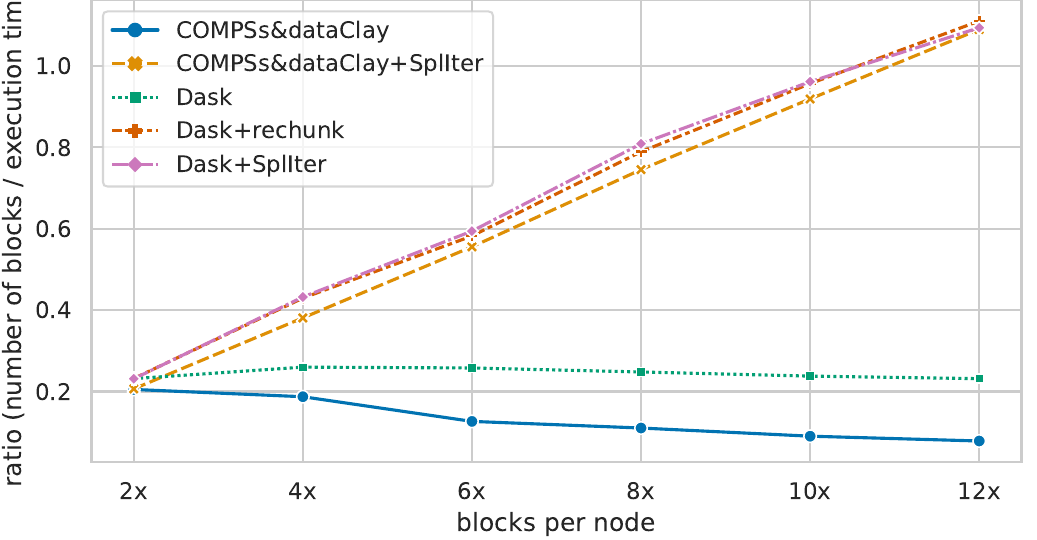}
    \caption{Evolution of the ratio number of blocks by execution time when increasing the number of blocks per node (higher is better).}
    \label{fig:kneighbors-grow}
\end{figure}

Figure~\ref{fig:kneighbors-grow} shows the ratio of number of blocks divided by the execution time. The plot starts at two blocks per core (the point where the baseline execution, the \emph{rechunk} and the \emph{SplIter} executions are all equivalent among them) and scales up to $12\times$ which is a total of 96 blocks for the training dataset.

We see how the Dask baseline is flat, meaning that the speed at which blocks are processed depends proportionally on the training dataset size. When the training dataset size is doubled, so is the execution time. We can see that the ratio is quite flat across all the execution meaning that this trend is present from 2 blocks per node up to 12 blocks per node (in that last execution the execution is 6 times bigger than the first one, and thus the execution is also 6 times longer). COMPSs~\& dataClay execution shows a bit more degradation, but follows the same general trend. This means that there is some additional overhead (as we have already discussed, due to leveragin in-memory intermediate data structures) but the execution times is proportional to the training dataset size too.

On the other hand, the ratio for all other executions increases linearly. The speed at which the \emph{SplIter} execution processes blocks is not proportional on the training dataset size, but better. When the training dataset size is doubled, the execution time is much less than twice. The root cause is the internal lookup trees. The \emph{kneighbors} stage is using the lookup trees generated during training; lookup operations on those trees are not $O(n)$ (linear) but $O(\log{n})$. The complexity of lookup trees explain the good behavior of \emph{SplIter} and \emph{rechunk}: duplicating the size of the tree does not duplicate the lookup time.

To summarize this comparison: generating trees directly from the block data results in a linear behavior while consolidating blocks (either with \emph{SplIter} or \emph{rechunk}) results in executions that are able to perform closer to the theoretical logarithmic complexity.

\subsubsection{Insights}

This application shows how addressing the fragmentation and granularity issues of the data can give benefits that go beyond serialization and scheduler overhead. For this specific application, we started from a $O(\log{n})$ theoretical complexity (the tree lookup stage) but observed it degrading into a $O(n)$ due to blocking (the baselines). Thanks to the use of either \emph{SplIter} or \emph{rechunk} we were able to greatly improve the performance --the execution time approaches once again the $O(\log{n})$.

This improvement requires an understanding of the algorithm. At this moment, \emph{rechunk} has no direct semantic to relate the size to the computing resources or number of nodes, so using it requires knowledge on the runtime computing resources. The \emph{SplIter} has the semantics related to data locality and computing resources, which results in a simpler programming model interface which is able to reach the same performance benefits as the \emph{rechunk} while avoiding data transfers.

\subsection{rechunk vs SplIter: discussion}

The \emph{rechunk} is an existing powerful tool for addressing inadequate granularity.  \emph{SplIter} is our proposal, which is an alternative that addresses inadequate granularity without data transfers nor data transformations.

In all the scenarios where data transfers and/or data transformations are expensive, our proposal yields better performance than using a \emph{rechunk} mechanism. Huge datasets and memory-bound executions are examples of such scenarios. On the other hand, materializing the structures  --i.e., doing a \emph{rechunk} or alternatively materializing the \emph{SplIter} partitions-- is favorable when the execution is compute-bound (due to small data structures and/or a relevant level of data reuse).

Another relevant aspect is the memory implications of the chunk size --an aspect that Dask documentation\cite{daskweb} highlights when advising how to ``select a good chunk size''. The chunks should be small enough so many of them fit in memory and the tasks do not run out of memory; the \emph{SplIter} breaks the dependency between task length and block size and solves this issue.

The overhead of using \emph{SplIter} in non-optimal scenarios is low (as shown in the Cascade SVM application), while the performance degradation of the \emph{rechunk} can be enormous (as in the Histogram application). Using the \emph{SplIter} is a more or less safe insurance in situations where knowing the optimal chunk size is impractical or flat out impossible.

\section{Conclusions}

\label{sec:conclusions}

In this article we have discussed and shown the \emph{SplIter} proposal, our contribution to improve iteration on distributed datasets in task-based programming models. At its core, the \emph{SplIter} is able to leverage iteration optimizations and data locality with minimal impact on programmability.

The evaluation has shown the behavior of the \emph{SplIter} across a variety of scientific applications, from several science domains (iterative or non-iterative applications, memory intensive applications, CPU intensive applications, machine learning, data analytics, etc.). The \emph{SplIter} is able to reduce the performance sensitivity to the block size. Given that the application developer may not know the computing environment, decoupling the application performance from the block size is a huge benefit in terms of programmability\cite{perez2008dependency} and performance portability\cite{kaiser2014hpx,aumage2021task} from the programming model perspective.

The \emph{SplIter} has been evaluated upon two different frameworks: COMPSs~\& dataClay and Dask. In general, the main idea behind the \emph{SplIter} could be applied to any task-based programming model (more generally: to any programming model with direct access to the blocks and the iteration code structures). The evaluation has shown that the \emph{SplIter} is able to compete and (in most situations) outperform the Dask \emph{rechunk}. 

We observed an application where \emph{rechunk} outperformed the \emph{SplIter} due to its very compute-intensive nature; this suggests that a new extension of the \emph{SplIter} may consist on providing materialized partitions. A materialized partition would be a new data structure, generated with memory copies but without inter-node transfers. For iterative algorithms that have a high computation to data ratio a materialized partition should be able to achieve same execution speeds as the \emph{rechunk} while avoiding all the inter-worker data transfers. Still, the performance difference between \emph{rechunk} and \emph{SplIter} was minor.

Complex applications can have multiple stages and iterations. In certain scenarios, the benefit of using \emph{SplIter} can also be observed at an algorithmic level: the quantity and shape of intermediate results may depend on the block size and that shape may have an important performance impact --we have seen this behavior in the \emph{k}-Nearest Neighbors, but having some kind of intermediate data structures is not an exclusive trait of this application. Using the \emph{SplIter} allows the application developer to exert their (domain-specific) expertise and greatly improve the performance in a portable way, without requiring prior knowledge from distribution techniques nor any insight of the hardware infrastructure topology.

\section{Acknowledgements}

This work has been funded by the European Commission's Horizon 2020 Framework program and  the European High Performance Computing Joint Undertaking (JU) under grant agreement No 955558 and by MCIN/AEI/10.13039/501100011033 and the European Union NextGenerationEU/PRTR (PCI2021-121957). Anna Queralt is a Serra H\'unter Fellow.

 \bibliographystyle{elsarticle-num} 
 \bibliography{cas-refs}

\end{document}